\documentclass[a4paper,twocolumn]{autartaccepted}

\usepackage{graphicx,psfrag,enumerate,subfigure}
\usepackage{amssymb}
\usepackage{amsmath}
\usepackage{calrsfs}

\newtheorem{definition}{Definition}
\newtheorem{theorem}{Theorem}

\newtheorem{lemma}{Lemma} 
\newtheorem{corollary}{Corollary} 
\newtheorem{proposition}{Proposition} 

\newtheorem{assumption}{Assumption}

\usepackage{color}
\usepackage[normalem]{ulem}
\usepackage{soul} 
\soulregister\cite7
\soulregister\ref7
\soulregister\pageref7

\begin{document}

\begin{frontmatter}

\title{Controller-jammer game models of Denial of Service 
  in control systems 
  operating over packet-dropping links} 

\thanks[footnoteinfo]{This work was supported by the Australian Research
  Council under 
  Discovery Projects funding scheme (project DP120102152)
  and, in parts, by
  US Air Force Office of Scientific Research (AFOSR) under grant number
  MURI FA 9550-10-1-0573, and the US National Science Foundation under
  award \#1151076.}

\author[First]{V.~Ugrinovskii}\ead{v.ugrinovskii@gmail.com}\and 
\author[Second]{C. Langbort}\ead{langbort@illinois.edu}

\address[First]{School of Engineering and IT, University of NSW
  at the Australian Defence Force Academy, Canberra, ACT, 2600, Australia.}
\address[Second]{Department of
Aerospace Engineering, University of Illinois at Urbana-Champaign.}


\begin{abstract}
The paper introduces a class of zero-sum games between the adversary and
controller as a scenario for a `denial of service' in a networked
control system. The communication link is modeled as a set of
transmission regimes controlled by a strategic jammer whose intention is
to wage an attack on the plant by choosing a most damaging regime-switching
strategy. We demonstrate that even in the one-step case, the introduced
games admit a saddle-point equilibrium, at which the jammer's
optimal policy is to randomize in a region of the plant's state space, thus
requiring the controller to undertake a nontrivial response which is
different from what one would expect in a standard stochastic control
problem over a packet dropping link. The paper derives conditions for
the introduced games to have such a saddle-point equilibrium. Furthermore,
we show that in more general multi-stage games, these conditions provide
`greedy' jamming strategies for the adversary.  
\end{abstract}

\end{frontmatter}

\section{Introduction and Motivation}\label{Intro}

The topic of control over a communication link has been extensively
studied in the past decade, with issues such as the minimum data rate for
stabilization \cite{Tak,Nair} and optimal quadratic 
closed-loop performance \cite{Sche,Imer} being the main focus. Other
issues of interest concern effects of channel-induced packet
loss and/or time-varying delays on closed-loop performance.

The majority of papers concerned with control over networks regards the 
mechanism of information loss in the network as probabilistic but
not strategic. In contrast, in the \emph{adversarial} networked control
problem, the communication link is controlled by a strategic
jammer who actively modifies the link to disrupt the control
goal. We broadly refer to such modifications/disruptions as Denial of
Service attacks; cf.~\cite{PIK-2011}.

A natural way to describe the adversarial networked control problem
is to employ a game-theoretic formulation. While the most immediate purpose
of the game-theoretic analysis may be to devise the best defence against
strategic attacks, it can also be used to predict possible attack
strategies. 
Originally proposed in~\cite{Amin}, the game-theoretic formulation of
adversarial networked control has been followed upon in a number of recent
papers including~\cite{Abhi,LaU3a}.  
A zero-sum dynamic game between a controller performing a finite horizon
linear-quadratic control task and a jammer, proposed in \cite{Abhi},
specifically accounted for the jammer's strategic intentions and
limited actuation capabilities. A startling conclusion of \cite{Abhi} was that
in order to maximally disrupt the control task, the jammer had to act in a
markedly different way than a legitimate,  non-malicious, packet-dropping
channel. More precisely, the jammer's saddle-point strategy was to
deterministically drop packets whenever the plant state was crossing
certain thresholds. Once this deterministic behavior is observed by the
controller, it can establish with certainty that an attack has taken place.  

The chief motivation behind the present work is to investigate whether it
is possible in principle for the attacker to be able to conceal its actions
by disguising as a packet dropping link. Naturally, this implies that the
controller has no prior information of the attack taking place.  
In~\cite{LaU3a}, to demonstrate such a possibility, we introduced a model
of adversarial networked control (ANC), which, while capturing the same
fundamental aspects of the problem as in \cite{Abhi}, modified the 
jammer's action space so that each jammer's decision corresponded to a choice
of a binary automaton governing the transmission rather than to passing/blocking
transmission directly. The  
corresponding one-step zero-sum game was shown to have a unique
saddle point in the space of mixed jammer's strategies. 
The optimal jammer's strategy was shown to randomly choose between two
automata, each having a nonzero probability to be selected. In turn, the
controller's best response to 
the jammer's  optimal strategy was to act as if it was operating
over a packet-dropping channel whose statistical characteristics were controlled
by the jammer. Since under normal circumstances the controller cannot be
aware of these characteristics, and cannot implement such a
best response strategy, the system performance is likely to be adversely
affected when the jammer follows its optimal strategy; see 
Section~\ref{LQR.reward.section}.  

In this paper, we show that such a situation is not specific to the
ANC problem considered in~\cite{LaU3a}, and it arises in a much more general 
zero-sum stochastic game setting. The only common
feature between our problem formulation and that in~\cite{LaU3a} is the
general mechanism of decision making adopted by the jammer. All other
attributes of the problem (the plant model, the assumptions on the
stage cost, etc.) are substantially more general, to the extent 
that unlike~\cite{LaU3a}, the saddle point cannot be computed
directly. Instead, for the one-step game, we obtain sufficient conditions
under which optimal jammer's strategies in the class of mixed strategies 
are to randomly choose between two actions. That is, to make a maximum
impact on the control performance, 
the jammer must act randomly, in contrast to~\cite{Abhi}.  

Our conditions for the one-step game are quite general, they apply to nonlinear
systems and draw on standard convexity/coercivity properties of payoff
functions. Under additional smoothness conditions, these conditions are
also necessary and sufficient. Also, we specialize these conditions
to three linear-quadratic control problems over a packet-dropping link. In
two of these problems, our conditions allow for a direct 
characterization of a set of plant's initial states for which optimal
jammer's randomized strategies exist. We also compute controller's optimal
responses to those strategies, which turn out to be nonlinear. The third
example revisits the problem setting in \cite{Abhi}, showing that our
conditions naturally rule out randomized jammer's behaviour in that
problem.

Our analysis of the one-step  game  can be thought of as reflecting a
more general 
situation where one is dealing with a one-step
Hamilton-Jacobi-Bellman-Isaacs (HJBI) min-max problem associated 
with a multi-step ANC problem.
Also, even the one-step formulation provides a rich insight into
a possible scenario of attacks on controller networks. For instance,
dynamic multistep jamming attacks can be planned so that at each step the
jammer chooses its actions based on the 
proposed formulation. Greedy jamming strategies where at each
time step the jammer pursues a strategy which is optimal only at this
particular time are discussed 
in Section~\ref{multistage}. 

Compared with the conference version~\cite{LaU3c}, here we have obtained a
new sufficient condition for the one-step ANC game to have a
nontrivial equilibrium. This condition does not
require the stage cost to be a smooth function, which potentially makes
this result applicable to the mentioned one-step HJBI min-max problems
where in general 
the smoothness of the value function cannot be guaranteed in advance; see
Theorem~\ref{exist_theo}. Also, the paper introduces a multi-stage game
model to study intelligent jamming attacks on linear
control systems. 

The paper is organized as follows. A general controller-jammer ANC game and
its connections with
intelligent jamming models are 
presented in Section~\ref{sec_mod}. The  conditions for the one-step ANC
game to have a non-pure saddle point are presented in
Section~\ref{Static.ANC.Section}. First we  
present a general sufficient condition suitable for analysis of 
general multi-input networked control systems. We then show that in the case of single-input
systems they are in fact necessary and sufficient (under an additional
smoothness assumption). Next, in 
Section~\ref{Examples} we demonstrate applications of these results to three
linear-quadratic static problems. In one of
these problems, which is an extension of the problem in~\cite{LaU3a}, the
jammer is offered an additional reward for undertaking actions concealing its
presence. In another problem, the jammer's actions take into account the cost
the controller must pay to mitigate the jammer's presence should the jammer
reveal itself. The third problem revisits the problem
in~\cite{Abhi}. Section~\ref{multistage} generalizes some of the results of
Section~\ref{Static.ANC.Section} about the
existence of saddle-point strategies to the case of multistage ANC games;
this generalization requires the plant to be linear. The Appendix
(Section~\ref{Appendix}) contains proofs of the results. 
Conclusions are given in Section~\ref{Conclusions}. 
\vspace{-20pt}

\paragraph*{Notation}
$\mathbb{R}^n$ is the $n$-dimensional Euclidean space of real vectors, 
$\mathbb{R}^n_+$ is the cone in $\mathbb{R}^n$ consisting of vectors whose
all components are non-negative.   
The unit simplex in $\mathbb{R}^N$ is denoted $\mathcal{S}_{N-1}$; i.e.,
$\mathcal{S}_{N-1}=\{p\in \mathbb{R}^N: 0\le p_j\le 1,
\sum_{j=1}^Np_j=1\}$.  $|\mathcal{F}|$ is
the cardinality of a finite set 
$\mathcal{F}$. $\delta_{i,j}$ is the Kronecker symbol, i.e.,
$\delta_{i,j}=1$ if $i=j$, otherwise $\delta_{i,j}=0$. 
For two sets $M$, $U$, $M\backslash U=\{x\in M\colon x\not\in
U\}$. The symbol $\mathrm{Pr}(\cdot)$ denotes probability of an event, and
$\mathrm{Pr}(A|z)$ denotes the conditional probability of an event $A$
given $z$. 

\section{Adversarial Network Control Games}
\label{sec_mod}

\subsection{The general system setup}\label{setup}
We consider a general setup, within which the evolution of a plant controlled over a
communication link subject to adversarial interference is described by  a
mapping $\mathbb{R}^n\times \mathbb{R}^m\to \mathbb{R}^n$,
\begin{equation}
\label{plant.t}
x^{t+1}=F_t(x^t,v^t),
\end{equation}
describing the response of the plant, when it is at state $x^t$, to an
actuator signal $v^t$ at time $t$. The actuator signal is determined by the
controller who transmits control packets over a randomly varying
communication link. 

The link can operate in one of several transmission regimes, and randomly
switches between them. The mechanism of the regime switching is controlled
by a strategic jammer, unbeknownst to the controller. For  
simplicity, we identify the set of transmission regimes with the set of
integers, so that $\mathcal{F}=\{1,\ldots,|\mathcal{F}|\}$. Once the link
updates its transmission regime, it randomly assumes either the passing or
blocking state. Thus, the transmission state of the communication
link is a binary random variable $b_t$, taking values 0 and 1 associated
with blocking and passing control packets, and $v^t=b^tu^t$.

Evolution of the  system over a time interval $\{0,\ldots,T\}$ is
captured by a bivariate stochastic process $\{(x^t,s^t)\}_{t=0}^T$, with
the state space $\mathcal{X}=\mathbb{R}^n\times \mathcal{F}$. We now
describe the dynamics of this bivariate process.

\textit{The initial state. }
Initially, at time $t=0$ the plant is at a state $x^0=x_0$, and the initial 
transmission regime of the communication link is $s^0=s$. It is assumed that at 
the time when the controller and the attacker make their decisions, both of
them know the current state of the plant and the transmission regime of the
link. In accordance with this assumption, they know $x^0$, $s^0$. 

\textit{State updates. } The system state  and the transmission regime of
the communication link are updated at every time $t$, in response
to controller's and jammer's actions, as described below.   

\emph{Controller actions. }
At every time instant $t=0,\ldots,T-1$, the controller observes the current
system state $x^t$ and the transmission regime $s^t$ of the communication
link. Based on this information, it generates a control input
$u^t\in \mathbf{R}^m$ and sends it via
the communication link. 

\emph{Jammer actions.} At every time step $t=0,\ldots, T-1$, the jammer observes
the plant state $x^t$, the transmission regime of the communication link
$s^t$, and the control signal $u^t$. Using this information, the jammer
selects a matrix from a predefined finite set of row stochastic matrices
$\mathcal{P}^t=\{P^t(a)\in \mathbb{R}^{|\mathcal{F}|\times
  |\mathcal{F}|}\colon a\in \mathcal{A}=\{1,\ldots,N\}\}$. That is, its
action $a^t$ is to draw a matrix $P^t(a^t)\in\mathcal{P}^t$. 

\emph{Communication link update. }
After the jammer has chosen its action $a^t=a$ and the corresponding
row stochastic matrix $P^t(a)$, the transmission regime of 
the link changes from $s^t$ to $s^{t+1}$ according to a Markov chain with
$\mathcal{F}$ as a state space and $P^t(a^t)=P^t(a)$ as the transition
probability 
matrix, i.e., 
\begin{eqnarray}\label{Pt}
\mathrm{Pr} (s^{t+1} = i| s^t=j, a^t=a)=P_{ji}^t(a) \quad
i,j\in \mathcal{F}.
\end{eqnarray}
The initial probability distribution of this Markov chain reflects that the
initial transmission regime is given to be $s^0=s$, i.e., 
$\mathrm{Pr}(s^0=i)=\delta_{i,s}$, $i\in\mathcal{F}$.

After the communication link randomly switches to the
transmission regime $s^{t+1}\in\mathcal{F}$, the transmission state of the
link is determined randomly and, given  $s^{t+1}$, it is conditionally
independent of the previous transmission regimes, the current and past
states of the plant and the controller's and jammer's actions. Thus, the probability
of the link to become passing is determined by a stochastic kernel
$\mathbf{q}^t$ on
$\{0,1\}$ given $\mathcal{F}$:  
\begin{eqnarray}
q_j^t = \mathbf{q}^t(b^t=1|j)&\triangleq& \mathrm{Pr}(b^t = 1 |s^{t+1}=j),
\quad j\in \mathcal{F}.  
\label{q_j}
\end{eqnarray}
We denote $q^t\triangleq (q_1^t, \ldots,q_{|\mathcal{F}|}^t)'$. It is worth
noting that the pair of processes $(s^{t+1},b^t)_{t=0}^{T-1}$ form a
controlled hidden Markov process, with $\mathcal{F}$ and $\{0,1\}$ as its 
the state space and the output alphabet, respectively, $\{\delta_{s^0,j},
j\in \mathcal{F}\}$ as its initial probability distribution, the sequence
of matrices $\{P^t(a^t)\}_{t=0}^{T-1}$ as the sequence of state
transition probability matrices, and the the sequence of stochastic kernels
$\{\mathbf{q}^t\}_{t=0}^{T-1}$ on $\{0,1\}$ given $\mathcal{F}$ as the
output probabilities. This hidden Markov process is denoted $\mathcal{M}$.       

\emph{Plant state updates.}
Next, the plant state is updated according to (\ref{plant.t}), in response to
control $u^t$ and the jammer's action $a^t$. Specifically, 
the transmission state of the link between the controller and the plant is
determined by the binary random variable $b^t$, whose
probability distribution given 
$(x^t,s^t)$, $u^t$ and $a^t$
can now be expressed as
\begin{eqnarray}
&&\mathrm{Pr}(b^t=1|s^t\!=j,a^t\!=a,x^t\!=x,u^t\!=u) 
=(P^t(a)q^t)_j, \quad \label{Prob.b=1}\\
&&\mathrm{Pr}(b^t=0|s^t=j,a^t=a) 
=1-(P^t(c,a)q^t)_j.\label{Prob.b=0}
\end{eqnarray}
Then the actuator signal $v^t$ becomes $v^t=b^tu^t$, and
according to (\ref{plant.t}), the plant's new state becomes $x^{t+1}$.
 
\begin{proposition}\label{Markov}
Given a sequence of controller's and jammer's actions, the joint state-link
process $\{x^t,s^t\}_{t=0}^T$ is a Markov process.  
\end{proposition}

The proof of this and all subsequent propositions is deferred to the
Appendix.     
 
The transition probability measure of the Markov process
$\{x^t,s^t\}_{t=0}^T$ defines a
sequence of stochastic kernels on $\mathcal{X}$, given
$u\in\mathbb{R}^m$, and $a\in\mathcal{A}$,
\begin{eqnarray}
  \label{Q.ker}
\lefteqn{Q_{t+1}(\Lambda\times S|x,j,u,a)} && \nonumber \\
&=& \mathrm{Pr}(x^{t+1}\in\Lambda,s^{t+1}\in S| x^t=x,s^t=j,u^t=u,a^t=a) \nonumber \\
&=&
\sum_{i\in S}P^t_{ji}(a)\left[q^t_i\chi_\Lambda(F_t(x,u)) +
  (1-q^t_i)\chi_\Lambda(F_t(x,0))\right] 
\end{eqnarray}
for $t=0, \ldots, T-1$; here $\Lambda$, $S$ are Borel subsets of
$\mathbb{R}^n$, $\mathcal{F}$ respectively, and $\chi_\Lambda(\cdot)$ is
the indicator function of the set $\Lambda$. 
The expectation with respect to the kernel $Q_t(\cdot |x,s,u,a)$
will be denoted $\mathbb{E}_t[\cdot|x,s,u,a]$.

\subsection{An adversarial network control game}
\label{sec_form}

We now consider a zero-sum stochastic game associated with the described setup. Its main components are:
\begin{itemize}
\item
The set $\mathcal{X}=\mathbb{R}^n\times \mathcal{F}$ as the state space of
the game; 
\item
the sets $\mathbb{R}^m$ and $\mathcal{A}$ as action spaces for the
controller and the jammer, respectively.
\item
The sequence of stochastic kernels (\ref{Q.ker}) on $\mathcal{X}$ given
$\mathcal{X}\times \mathbb{R}^m\times \mathcal{A}$ and 
the Markov process $\{(x^t,s^t)\}_{t=0}^T$ associated with 
given sequences of controller's and jammer's actions.   

\item
A real-valued finite horizon cost 
\begin{equation}
\label{cost_cont}
\Upsilon\left(\{x^t\}_{t=0}^T, \{u^t\}_{t=0}^{T-1} \right) = \sum_{t=0}^{T-1}
\sigma^t(x^t,u^t) + \sigma^T(x^T). 
\end{equation}
The stage cost $\sigma^t(x^t,u^t)$ in (\ref{cost_cont}) 
symbolizes the performance loss incurred by the system (1) when the control
$u^t$ is applied to the system while it is in state $x^t$. In the
absence of the jammer, the controller would be expected to minimize this
loss.

\item
A real-valued function representing the cost of jammer's actions
$a^t$, $t=0,\ldots, T$:  
\begin{equation}
\label{cost_jam}
\Gamma(\{a^t\}_{t=0}^{T-1},\{s^t\}_{t=0}^{T-1}) =\sum_{t=0}^{T-1}
g^t(a^t,s^t). 
\end{equation}
It describes how the jammer's resources and
  incentives are affected, when it takes its actions to control 
  the communication link. The stage cost $g^t(a^t,s^t)$ in (\ref{cost_jam}) 
symbolizes the cost incurred by the jammer when at time $t$ it selects
$P^t(a^t)$ as a state transition probability matrix for the Markov
process $\mathcal{M}$. Normally, the jammer
  would be expected to minimize this cost.  

\item
A total cost of the controller's and jammer's actions
\begin{eqnarray}
\lefteqn{\Sigma\left(\{x^t\}_{t=0}^T, \{u^t\}_{t=0}^{T-1}, 
    \{a^t\}_{t=0}^{T-1}\right)} && \nonumber \\
&&=\Upsilon\left(\{x^t\}_{t=0}^T, \{u^t\}_{t=0}^{T-1} \right)
-\Gamma\left(\{a^t\}_{t=0}^{T-1},\{s^{t-1}\}_{t=0}^T\right). \qquad
\label{cost_total}
\end{eqnarray}
The minus sign in (\ref{cost_total}) indicates that degrading the control performance comes at a cost
  to the jammer, who must spend its resources to achieve its goals.

\end{itemize}

In what follows we consider a class of control input sequences
$\{u^t\}_{t=0}^T$ and a class of jammer's strategies consisting of sequences
$\{p^t\}_{t=0}^{T-1}$ of probability vectors $p^t$ over $\mathcal{A}$. 
Associated with every such
sequence of the controller's and jammer's strategies
$\{u^t\}_{t=0}^{T-1}$, $\{p^t\}_{t=0}^{T-1}$ and system and link
initial conditions $x^0,s^0$ is a 
probability distribution on the space of sequences of system and
communication link states endowed with the $\sigma$-algebra of all Borel
subsets\footnote{In (\ref{total.Q}), $x_t$ is an arbitrary vector in
  $\mathbb{R}^n$, in contrast to the state $x^t$ of the plant at time
  $t$. Likewise, $s_t$ is an
  integer in $\mathcal{F}$, in contrast to the random state $s^t$ of the
  Markov process $\mathcal{M}$ with values in $\mathcal{F}$.}      
\begin{eqnarray}
  \label{total.Q}
\lefteqn{\mathbf{Q}^{u,p}(dx_1\ldots,dx_T,s_1,\ldots,s_T|x_0,s_0)}
&& \nonumber \\
&&= 
Q^{u,p}_1\!(dx_1,\!s_1|x_0,s_0) \ldots  
Q^{u,p}_T(dx_T,\!s_T|x_{T-1},\!s_{T-1}),
\end{eqnarray}
generated by the sequence of stochastic kernels 
\begin{eqnarray}
  \label{Q.ker.p}
\lefteqn{Q^{u,p}_{t+1}(dx,s|x,j)=
\sum_{a\in\mathcal{A}} p^t_aQ_{t+1}(dx,s|x,j,u^t,a),}  
\end{eqnarray}
where $Q_t$ is the stochastic kernel defined in (\ref{Q.ker}).
The expectation with respect to the probability measure
$\mathbf{Q}^{u,p}$ will be denoted $\mathbb{E}^{u,p}$. 
We will omit the superscripts $^{u,p}$ when this does not lead to a
confusion.   

\begin{definition}
The zero-sum stochastic game with the state space $\mathcal{X}$, action
spaces $\mathbb{R}^m$ for the controller and $\mathcal{A}$ for the jammer,
the stochastic kernels $\{Q^t,t=0,\ldots,T\}$ defined in (\ref{Q.ker}) and the 
payoff function $\Sigma$ is called the adversarial networked control (ANC) game
associated with the Markov process $\{(x^t,s^t)\}_{t=0}^T$ and costs
(\ref{cost_cont}), (\ref{cost_jam}), (\ref{cost_total}). 
For a given plant initial condition $x^0=x$ and an initial transmission
regime of the link
$s^0=s$, solving the 
ANC game requires one to find (if they exist) strategies for the jammer
$p^*=\{(p^*)^t\}_{t=0}^{T-1}$ and 
for the controller $u^*=\{(u^*)^t\}_{t=0}^{T-1}$ 
such that
\begin{eqnarray}
 \lefteqn{\mathbb{E}^{u,p^*}\left[\Sigma\left(\{x^t\}_{t=0}^T, 
  \{(u^*)^t\}_{t=0}^{T-1},  \{a^t\}_{t=0}^{T-1}\right)\right]} && \nonumber \\
&& 
=\inf_{\{u^t\}_{t=0}^{T-1}}\sup_{\{p^t\}_{t=0}^{T-1}}
 \mathbb{E}^{u,p}\left[\Sigma\left(\{x^t\}_{t=0}^T, \{u^t\}_{t=0}^{T-1},
    \{a^t\}_{t=0}^{T-1}\right)\right] \nonumber \\
&& 
=\sup_{\{p^t\}_{t=0}^{T-1}}\inf_{\{u^t\}_{t=0}^{T-1}}
\mathbb{E}^{u,p}\left[\Sigma\left(\{x^t\}_{t=0}^T, \{u^t\}_{t=0}^{T-1},
    \{a^t\}_{t=0}^{T-1}\right)\right].\nonumber \\
  \label{equi-point.t} 
\end{eqnarray}
\end{definition}

\subsection{Additional comments and special cases}

\subsubsection{Relation to conventional DoS attacks on
  communications}\label{relation}

In the model presented in Section~\ref{setup}, the adversary 
controls information transmission strategically and indirectly, unbeknownst
to the controller. The motivation for introducing the indirect jamming 
model is to reduce the possibility of attack detection by adding intelligence to
jamming strategy and attempting to force the controller to believe that
performance degradation is due to poor link conditions and not due to the
presence of the attack, i.e., our model is consistent with objectives of
\emph{intelligent Denial-of-Service
  jamming}~\cite{PIK-2011}. We now present two intelligent jamming 
scenarios which our model captures.

\begin{figure}
  \centering
\includegraphics[width=0.9\columnwidth]{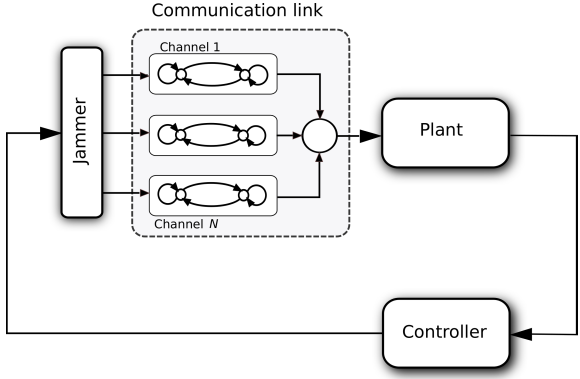}
  \caption{An attack by re-routing control packets. The jammer
    strategically selects a binary channel which randomly blocks or passes
    transmission of control information from the controller to the plant.}
  \label{fig:system}
\end{figure}

\emph{An attack by re-routing. }
In this scenario, the controller and the plant are connected by a multi-hop
network (i.e., a graph), with each transmission regime corresponding to a
path connecting the controller and the plant vertices in that graph. Assuming that the graph is acyclic, the set of
such paths has finite (yet possibly large) cardinality, so it can be modeled by
our set $\mathcal{F}$.  
When a jammer intercepts a message at a vertex and re-routes it on an
outgoing edge of its choosing it, essentially, modifies the characteristics
(packet-drop rate, delay, etc.) of the communication channel experienced by
the controller.  It can thus, in effect, be thought of as controlling the
characteristics of a communication channel or as `switching' channels.  
Accordingly, the jammer can be thought of having the action set
$\mathcal{A}=\mathcal{F}$, and the row stochastic matrices (\ref{Pt})
describing changes of the transmission regimes in this example are defined as
\begin{equation}
  \label{Pt.1}
  P^t_{ji}(a)=\begin{cases}1, & \text{if $i=a$},\\
                                 0, & \text{if $i\neq a$},
                               \end{cases}
\end{equation}
i.e., the control information is transmitted through the channel selected
by the jammer. 

The controller may know the previous channel
used for transmission (e.g., from an acknowledgement from the plant), and
therefore in general its decision $u^t$ can depend on $s^t$. However, even
though the controller has knowledge about the previously used channel and
can use this knowledge to select a channel which it wants to use, the 
attacker overrules this selection and redirects control packets through a
different channel. The
actual mechanism through which (re)routing, and the ensuing `switching',
is performed in the network is not important for our purposes. All that
matters is that jammer's actions eventually affect the row stochastic
matrix describing transmission, as (\ref{Pt.1}) shows. In that
sense, the  picture in Fig.~\ref{fig:system} is a good metaphor (it is not
an exact technical description) of the jamming process.   

Letting the jammer select the channel 
randomly, according to a probability vector
$p^t=(p^t_1,\ldots,p^t_N)'$ over $\mathcal{F}$  
will allow it to control the probability of packet dropouts, since in this
case 
\begin{eqnarray}
\label{bt=1}
\mathrm{Pr}(b^t=1|s^t=j)=(p^t)'q^t.
\end{eqnarray}
As a result, the end-to-end statistical characteristics of
the communication link depend on the jammer's selection of the probability
distribution $p^t$ but are independent of the past channel 
$s^t$ known to the controller. Hence, such characteristics are
difficult for the controller to predict. Implications of this observation
will be considered in detail in the next sections.

\emph{An attack by modifying channel characteristics. } 
Where there is a single physical channel between the
controller and the plant, the adversary can interfere with transmission by
means of intercepting control packets and `filtering' them through one of
several binary automata available to it. Each `filter'-automaton determines
the transmission regime in this case. Indeed, let again
$\mathcal{A}=\mathcal{F}$ and let $P^t_{ji}(a)$ be defined as in
(\ref{Pt.1}). Then the probability for the link to become passing when
the jammer selects automaton $a^t=i$ becomes the function of the jammer's
action, since according to~(\ref{Prob.b=1}) we have:   
\begin{eqnarray*}
\mathrm{Pr}(b^t=1|s^t=j,a^t=i)=q^t_i
=\mathrm{Pr}(b^t=1|a^t=s^{t+1}=i).
\end{eqnarray*}
Again, we see that this probability does not depend on the transmission
regime $s^t$ utilized at the previous time step but depends on the jammer's
action. As in the previous attack model considered, the end-to-end
statistical characteristics of the  communication link depend on the
jammer's actions and hence are  difficult for the controller to
predict. Letting the jammer  select its actions according to a probability
distribution $p^t=(p^t_1,\ldots,p^t_N)'$ will allow it to exercise even
greater control over the distribution of $b^t$, according to (\ref{bt=1}).

In this scenario, since the transmission regimes are fully controlled by
the jammer, the controller is unlikely to be able to observe past jammer's
actions. In view of this restriction, the controller's feasible strategies are
limited to state feedback strategies $u^t=u^t(x^t)$, which are a special
case of more general control strategies dependent on $s^t$
mentioned earlier. 

In both attack scenarios presented above, direct jamming has been
forsaken in favour of intelligently controlling the probability of control
packets delivery. As was mentioned, a direct control over 
transmission was shown in~\cite{Abhi} to result in a deterministically
predictable jamming strategy. We will show in this
paper that the proposed intelligent jamming strategies lead to a distinctly
different jammer behaviour while forcing the controller to respond in a manner
distinct from what one would expect in the situation where packet loss is
benign. 

\subsubsection{Strategy spaces}

In the ANC game (\ref{equi-point.t}),  
an optimal jammer's strategy is sought in the set of mixed strategies, i.e.,
probability distributions on $\mathcal{A}$. On the other hand,
the controller does not use mixed strategies. Indeed, in a number of control 
problems considering benign loss of communication,  optimal
control laws are sought in the class of linear non-randomized functions of 
the (estimated) state of the plant; e.g, see~\cite{Sche}. It is this sort of
situation that we consider as a target for stealthy intelligent jamming ---
unless 
the controller knows that the system is under attack, it has no reason to
deviate from its optimal policy.

\subsubsection{The actuation model and control information}
The proposed actuation model assumes that 
when the link drops control packets,
the actuator signal $v^t$ is set to 0. The `hold' situation where 
$v^t$ is set to the last received control signal
$u^-$ whenever $u^t$ is dropped, can be reduced to this case by the change
of control variables. 

Our control model includes feedback controls that use
the full information about the state of the underlying Markov process
$\mathcal{M}$. In practice, it may not always be possible for the
controller to observe this state; the second example in
Section~\ref{relation} alludes to such 
situation. Games with partial information are known to be much more
difficult to solve, and the existence of a saddle point is more difficult
to establish if possible at all. Considering a larger class of
controllers which includes Markov control laws allows us to circumvent this
difficulty, and obtain conditions under which the ANC  
game has an equilibrium (as will be shown, under these conditions, the
controller's 
optimal response is indeed Markov). The upper value of this game then
sets the lower bound on control performance which can be achieved when, for
instance, the controller is agnostic about the transmission characteristics
of the communication link.

\section{Static ANC games}\label{Static.ANC.Section}

\subsection{One-stage ANC games}

Our aim is to determine conditions under
which an ANC game admits non-pure saddle points. By non-pure saddle point
we mean a saddle point of the game (\ref{equi-point.t}) consisting of
sequences of non-pure strategy vectors $(p^*)^t$, each being a linear
combination of two or more vertices of the simplex
$\mathcal{S}_{N-1}$, and the corresponding controller's best
responses $(u^*)^t$. As demonstrated in Section~\ref{relation},
non-pure policies allow the attacker to
exercise an intelligent DoS jamming by randomizing its actions and
controlling the probability of control packet transmission. To facilitate characterization of non-pure ANC
saddle-point strategies, in this section we consider one-stage ANC games which can be
regarded as a special case of the ANC game (\ref{equi-point.t}) played over
a one time step horizon.  

To emphasize the one-stage nature of the game, we temporarily suppress the time variable $t$, and adopt a
simplified notation, where the variables current at time $t$ are not
indexed, while the variables which are updated and become current at time $t+1$
are marked with the superscript $^+$. With this convention, a one-step
version of the general model presented in Section~\ref{sec_mod} involves:  

\begin{itemize}
\item
A plant initial state $x$ and an initial transmission regime of the
communication link $s$, which captures characteristics of the link between
the controller and the plant before the jamming attack is undertaken.     

\item
Controller and jammer actions $u$ and $a$, with $a$ selected
according to a probability distribution $p$ over $\mathcal{A}$. 

\item
A transmission mode $s^+$ of the link, randomly drawn from $\mathcal{F}$
according to the probability distribution $P_{s,\cdot}(a)$ corresponding to
the $s$-th row of the matrix $P(a)$ selected by the jammer; see (\ref{Pt}).
 
\item
The probability of the link operating in the transmission regime
$s^+\in\mathcal{F}$ to become passing, determined by equation
(\ref{q_j}). The vector of these 
probabilities is denoted $q=(q_1, \ldots, q_{|\mathcal{F}|})'$.

\item
The binary random variable $b$ capturing the transmission state of the
communication link at the time of plant state
update. According to (\ref{Prob.b=1}), when the jammer's actions have
a distribution $p$, the probability of the link to become 
passing is 
$\sum_{a=1}^{N}p_a(P(c,a)q)_s$.

\item
The updated state of the plant $x^+$, which is the terminal plant state in the
one-stage ANC game. From (\ref{plant.t}), 
\begin{equation}
  \label{plant}
x^+=F(x,bu).  
\end{equation}
\item
The payoff function associated with the controller's and jammer's
actions and the given initial system state $x$ and the link transmission regime
$s$, 
\begin{equation}
\label{sigma}
\Sigma(x^+,u,a)=\sigma^0(x,u)+\sigma^1(x^+)-g^0(a,s). 
\end{equation}
The latter is a special case of the payoff (\ref{cost_total}). 
\end{itemize}

In the one-step case, the probability measure (\ref{total.Q}) reduces to
the conditional probability measure on 
$\mathcal{X}$ given that the plant starts at $x$ and the link
is initiated at an initial transmission regime $s$
\begin{eqnarray}
  \label{total.Q.0}
\mathbf{Q}^{u,p}(dx^+,s^+|x,s)&=& \sum_{a=1}^{N}
p_a P_{ss^+}(a)\left[q_{s^+}\delta(x^+-F(x,u)) \right. \nonumber \\
&&\hspace{-.5cm}+\left.(1-q_{s^+})\delta(x^+-F(x,0))\right]dx^+.
\end{eqnarray}

\begin{definition}\label{greedy.ANC.game}
The one-stage ANC game is to find (if they exist) a strategy for the jammer
$p^*\in\mathcal{S}_{N-1}$ and a feedback strategy for the
controller $u^*$ such that 
\begin{eqnarray}
  \label{equi-point.t.0}
\mathbb{E}^{u^*,p^*}\Sigma(x^+,u^*,a)=J_1=J_2,
\end{eqnarray}
where $J_1$ and $J_2$ are the upper and the lower values of the game,
\begin{eqnarray}
J_1&=&\inf_{u\in\mathbb{R}^m} \sup_{p\in \mathcal{S}_{N-1}}  \mathbb{E}^{u,p} 
\Sigma(x^+,u,a), \label{inf_sup} \\
J_2&=&\sup_{p\in \mathcal{S}_{N-1}} \inf_{u\in\mathbb{R}^m}  \mathbb{E}^{u,p} 
\Sigma(x^+,u,a).
\label{sup_inf}
\end{eqnarray}
The pair $(u^*,p^*)$ is then a saddle point of the game.
\end{definition}

We wish to determine whether the ANC game has non-pure
saddle points. As was explained earlier, such
saddle points signify a possibility for the jammer to randomly choose
between several actions as its optimal strategy.

\subsection{Existence of a saddle point in one-stage ANC games}\label{main}

Our result on the existence of an equilibrium of
the one-stage game (\ref{equi-point.t.0}) relies on the following
assumption. 

\begin{assumption}\label{sigma.A3}
For every $x\in\mathbb{R}^n$,  $a\in \mathcal{A}$, the functions 
$\Sigma(F(x,0),\cdot,a)$ and $\Sigma(F(x,\cdot),\cdot,a)$ are continuous,
coercive\footnote{Recall that a function $h:\mathbb{R}^m\to \mathbb{R}$ is coercive
if there exists a function $\eta:\mathbb{R}_+\to \mathbb{R}$ such that
$\lim_{y\to+\infty} \eta(y)=+\infty$ and $h(u)\geq \eta(\|u\|)$ for all
$u\in \mathbb{R}^m$.} and convex functions $\mathbb{R}^m\to \mathbb{R}$.
\end{assumption}

According to (\ref{total.Q.0}), the
expected cost in (\ref{equi-point.t.0}) has the form 
\begin{eqnarray}
  \label{ESigma}
\mathbb{E}^{u,p} \Sigma(x^+,u,a)=p'h^{x,s}(u),
\end{eqnarray}
where $h^{x,s}(\cdot)$ is a vector function $\mathbb{R}^m\to
\mathbb{R}^{N}$, whose $i$-th component represents 
the conditional expected cost value given a jamming action $a=i$:
\begin{eqnarray}
  \label{hj}
h^{x,s}_i(u)&=& \mathbb{E}^{u,p}[\Sigma(x^+,u,a)|a=i] \nonumber \\ 
&=&(P(i)q)_s \Sigma(F(x,u),u,i)  \nonumber \\
&& \hspace{.5cm}+(1-(P(i)q)_s)\Sigma(F(x,0),u,i). 
\end{eqnarray}
Under Assumption~\ref{sigma.A3}, each function $h_i^{x,s}(\cdot)$ is continuous, convex and coercive for every
$s\in\mathcal{F}$ and $x\in\mathbb{R}^n$.

\begin{theorem}\label{BO.T4}
Under Assumption~\ref{sigma.A3}, for every initial pair $(x,s)$, the
one-stage ANC game has a finite value, i.e.,  
$
-\infty<J_1=J_2<\infty,
$
and the game has a (possibly non-unique) saddle point. Furthermore, for
every $x$, there exists a compact set $U(x)\subset \mathbb{R}^m$ which
contains the controller's optimal response $u^*=u^*(x,s)$ and is such that 
\begin{eqnarray}
J_1&=&\inf_{u\in U(x)} \sup_{p\in \mathcal{S}_{N-1}}  \mathbb{E}^{u,p} 
\Sigma(x^+,u,a)  \label{inf_U_sup} \\
&=&\sup_{p\in \mathcal{S}_{N-1}} \inf_{u\in U(x)} \mathbb{E}^{u,p} 
\Sigma(x^+,u,a) = J_2.
  \label{J1J2}
\end{eqnarray}
\end{theorem}

Unlike standard results on
the existence of a saddle point in a static zero-sum convex-concave
game~\cite{B&O}, the 
strategy space of the minimizer is not bounded in the game
(\ref{inf_sup}), (\ref{sup_inf}). Theorem~\ref{BO.T4} shows that the
minimization over $u\in\mathbb{R}^m$ in (\ref{inf_sup}) and 
(\ref{sup_inf}) can be reduced to a minimization over a compact set. 
According to (\ref{J1J2}), every saddle point of the static game played on
$U(x)\times \mathcal{S}_{N-1}$ is a saddle point of the game played on
$\mathbb{R}^m\times \mathcal{S}_{N-1}$.  

\subsection{General sufficient conditions for the existence of optimal
 non-pure jammer strategies}\label{main.1}

Theorem~\ref{BO.T4} asserts the existence of a saddle-point strategy for
the ANC game~(\ref{equi-point.t.0}). However, it does not answer the question
as to whether the jammer's corresponding strategy
is randomized or not, which is of central
importance for a possibility of intelligent jamming attacks. To answer this
question, we now provide sufficient and, under additional technical assumptions,
necessary and sufficient conditions for the existence of non-pure saddle
points in the ANC game~(\ref{equi-point.t.0}). These 
conditions characterize the controller-jammer games in which the jammer
randomizes its choice of optimal strategies. As we will see in
Section~\ref{Examples}, this will force the controller to respond in a
non-obvious manner in order to remain optimal.    

\begin{lemma}\label{suff_cond}
Let $H=[H_1,\ldots, H_N]'$ be a convex
vector-valued function $\mathbb{R}^m \rightarrow \mathbb{R}^{N} $ (i.e.,
each component of $H$ is a convex function). Assume there 
exist $u^* \in \mathbb{R}^m$ and a set $\mathcal{G} \subset \mathcal{A}$ with
$|\mathcal{G}| \geq 2$ such that  
\begin{enumerate}[(a)]
\item
$H_i(u^*) = \max_{1 \leq j \leq N} H_j(u^*)$ for all $i \in \mathcal{G}$.  
\item 
For all $u \neq u^*$, there exists $i \in \mathcal{G}$ such that $H_i(u) > H_i(u^*)$.
\item
$u^*$ is not a minimum of any $H_i$, $i \in \mathcal{G}$.
\end{enumerate}
Then, there exists a non-pure vector $p^* \in \mathcal{S}_{N-1}$, with
support $I( p^* ) \subseteq \mathcal{G}$ such that 
\begin{eqnarray*}
\inf_{u\in \mathbb{R}^m}\! \sup_{p \in \mathcal{S}_{N-1}}\! p' H(u)
&=& (p^*)' H(u^*) = \!\!\sup_{p \in \mathcal{S}_{N-1}}\!  \inf_{u\in
  \mathbb{R}^m}\! p' H(u).   
\end{eqnarray*}
\end{lemma}

Our first result about the  ANC game
(\ref{equi-point.t.0}) now follows.   

\begin{theorem}
  \label{Suff_theorem}
Suppose Assumption~\ref{sigma.A3} holds and for every $x\in\mathbb{R}^n$,
$s\in\mathcal{F}$, the function 
$
H(u)\triangleq h^{x,s}(u)=[h_1^{x,s}(u),\ldots,
h_{N}^{c,s}(u)]'
$
satisfies conditions (a)-(c) of Lemma~\ref{suff_cond}; i.e., there exist
$u^*$ and a set $\mathcal{G}\subset \mathcal{A}$ for which conditions
(a)-(c) hold. Then the static ANC game 
(\ref{equi-point.t.0}) has a non-pure saddle point $(u^*,p^*)$.  
\end{theorem}

Theorem~\ref{Suff_theorem} asserts the existence of a non-pure saddle
point in the 
one-stage game under consideration. 
The jammer can use Theorem~\ref{Suff_theorem} to assess whether  
it can launch a randomized jamming attack, as follows. 
The jammer
observes the state of the system $(x,s)$ and the control signal $u$, then
checks whether $u$ belongs to the set  
\begin{eqnarray*}
\mathcal{U}^{s,x}=\{u^*\colon \exists \mathcal{G}\subseteq\mathcal{A}, |\mathcal{G}|\ge 2,
\mbox{ s.t. $H(u^*)=h^{x,s}(u^*)$}\\\mbox{verifies
  conditions of Lemma~\ref{suff_cond}}\}.    
\end{eqnarray*}
If $u\in\mathcal{U}^{s,x}$ and the corresponding set $\mathcal{G}$ exists, the
jammer can launch an attack by randomly selecting $a\in \mathcal{G}$,
according to an 
arbitrary probability distribution $p^*$ supported on $\mathcal{G}$ (it 
is shown in the proof of Lemma~\ref{suff_cond} that any such vector can serve as
the jammer's best response; see 
Appendix). The controller also can carry out a similar 
analysis if it knows the state of the system $(x,s)$ and the set
$\mathcal{A}$, to determine whether the 
control input $u$ it has chosen can trigger a randomized jamming
behaviour. For this, the controller must verify whether
$u\in\mathcal{U}^{s,x}$. When the controller does not know the transmission
regime $s$, then a conservative test can be performed to check whether
$u\in\cup_s\mathcal{U}^{s,x}$.  
Neither the jammer nor the controller need to compute their optimal
strategies to perform this analysis. In the remainder of
this Section and in Section~\ref{Examples}, we will further particularize
conditions of Theorem~\ref{Suff_theorem}, to
provide a better insight into how Theorem~\ref{Suff_theorem} enables such
an analysis. 

Lemma \ref{suff_cond} and Theorem~\ref{Suff_theorem} make no
assumption regarding the smoothness of the function $H$, which is useful for
application to dynamic ANC games, where the game (\ref{inf_sup}),
(\ref{sup_inf}) may arise as an Isaacs equation from application of the
Dynamic Programming. The value function in such an equation 
may not be smooth (e.g., in~\cite{LaU3a} it was shown to be only
piece-wise smooth). However, Theorem~\ref{Suff_theorem} can be
sharpened when local differentiability holds. This will be
demonstrated in the next section. 

\subsection{Necessary and sufficient conditions for strategic
  jamming}\label{main.2} 

The analysis in this section requires control inputs $u$ to be one-dimensional,
and 
the functions $h_i^{x,s}(\cdot)$, $i\in \mathcal{G}$ must be piece-wise
smooth. Furthermore, we assume that the set $\mathcal{G}$ consists of only
two actions, $\mathcal{G}=\{a_1,a_2\}$.

\begin{assumption}\label{sigma.A1'}
Suppose the space of control inputs is one-dimensional. Also, for every $x\in
\mathbb{R}^n$, $a\in\mathcal{A}$, the functions
$\Sigma(F(x,\cdot),\cdot,a)$ and  
$\Sigma(F(x,0),\cdot,a)$ are convex functions defined on 
$\mathbb{R}^1$, which are continuously differentiable on $\mathbb{R}^1$,
perhaps with the exception of a finite number of 
points; let $U^d(x)$ be the set of such points. 
\end{assumption}

Clearly, under Assumption~\ref{sigma.A1'}, the functions $h_{a_1}^{x,s}(\cdot)$,
$h_{a_2}^{x,s}(\cdot)$ are convex. Also, these functions are continuously
differentiable at every $u$, except for $u\in U^d$.   

\begin{theorem}\label{saddle.point}
Suppose Assumption \ref{sigma.A1'} holds. 
\begin{enumerate}[(i)]
\item
If for every $x,s$, there exists $\bar u\not\in U^d$ such that 
\begin{eqnarray}\label{F=F}
&& h_{a_1}^{x,s}(\bar u)=h_{a_2}^{x,s}(\bar u), \\
\label{FF>F}
&&h_i^{x,s}(\bar u)\ge h_a^{x,s}(\bar u), \quad \forall i\in\mathcal{G},
a\not\in \mathcal{G}, 
\end{eqnarray}
and one of the following conditions hold: either
\begin{equation}\label{dF.dF<0}
\left(\frac{d h_{a_1}^{x,s}(\bar u)}{d u}\right)
\left(\frac{d h_{a_2}^{c,s}(\bar u)}{d u}\right)<0,
\end{equation}
or
\begin{equation}\label{dF=dF=0}
\frac{dh_{a_1}^{x,s}(\bar u)}{d u}=
\frac{dh_{a_2}^{x,s}( \bar u)}{d u}=0,
\end{equation}
then  the zero-sum game (\ref{inf_sup})
admits a non-pure saddle point $(u^*, p^*)$, $u^*\not\in U^d$, with $p^*$
supported on $\mathcal{G}=\{a_1,a_2\}$.
\item
Conversely, if the zero-sum game (\ref{inf_sup})
admits a non-pure saddle point $(u^*, p^*)$, $u^*\not\in U^d$, with $p^*$
supported on $\mathcal{G}=\{a_1,a_2\}$, then there
exists $\bar u\not\in U^d$ such that either (\ref{F=F}), (\ref{dF.dF<0}) or  
(\ref{F=F}), (\ref{dF=dF=0}) hold.
\end{enumerate}
\end{theorem}

\begin{figure}[t]
\psfrag{u2}{$\bar u$}
\psfrag{u1}{$u^0$}
\psfrag{h1}{$h_{a_1}^{x,s}$}
\psfrag{h2}{$h_{a_2}^{x,s}$}
  \centering
\subfigure[\label{(a)}]{%
\includegraphics[width=0.2\textwidth]{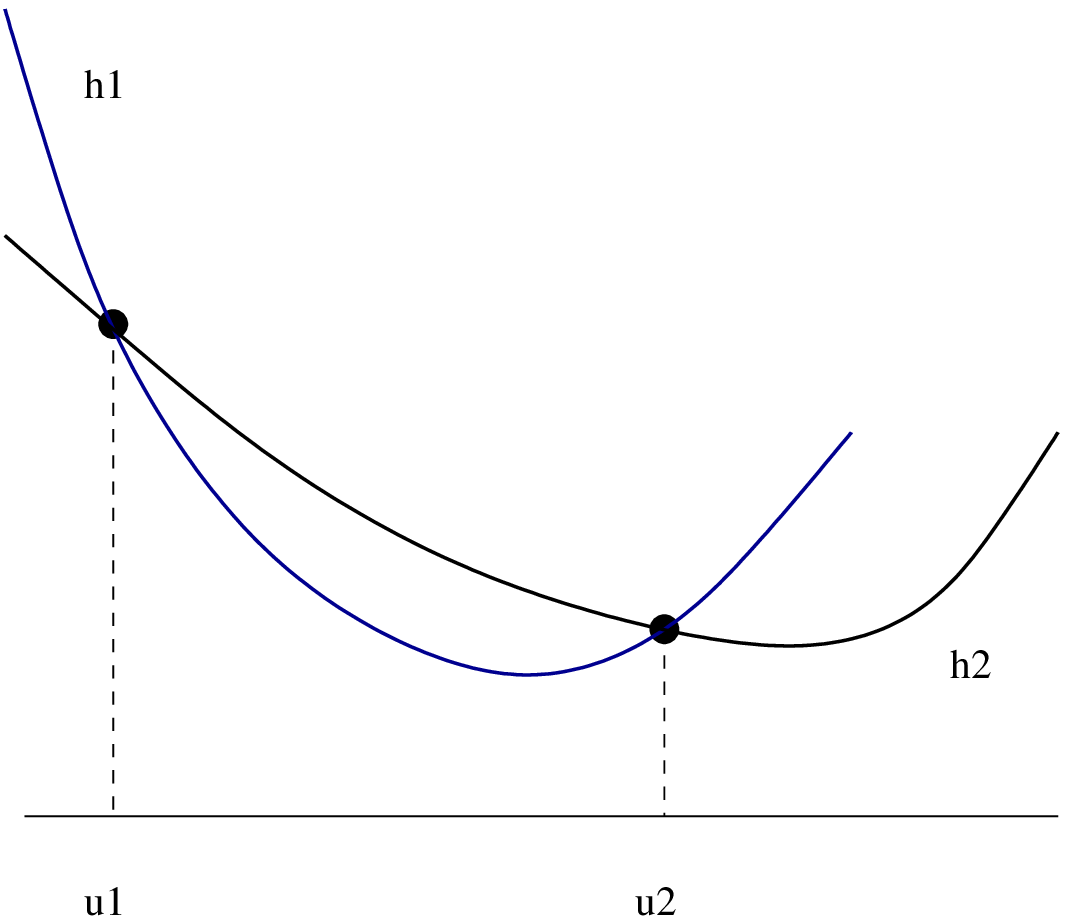}
}
\quad
\subfigure[\label{(b)}]{%
\includegraphics[width=0.2\textwidth]{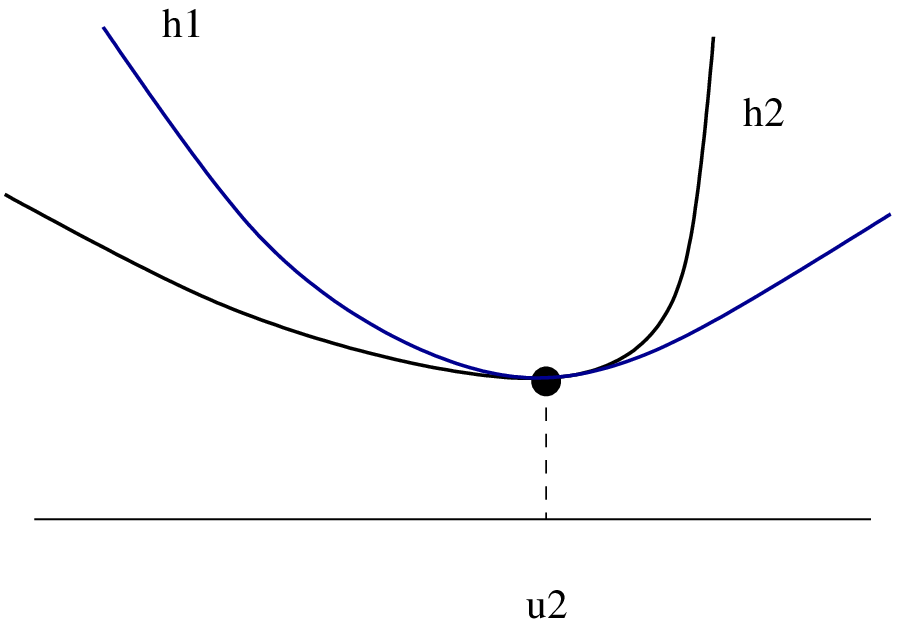}
}
  \caption{Graphs of the functions $h_{a_1}^{x,s}( \cdot), 
    h_{a_2}^{x,s}(\cdot)$ in 
    Theorem~\ref{saddle.point}. In Fig.~(a), $\bar u$ satisfies conditions
    (\ref{F=F}) and (\ref{dF.dF<0}), but $u^0$ does not satisfy
    (\ref{dF.dF<0}). In Fig.~(b), $\bar u$ satisfies both
    (\ref{F=F}) and (\ref{dF=dF=0}).}
  \label{h.figure}
\end{figure}

The conditions of Theorem~\ref{saddle.point} are illustrated in
Figure~\ref{h.figure}.
We note that
conditions (\ref{F=F}), (\ref{FF>F}) are a special case of 
condition~(a) of Theorem~\ref{Suff_theorem}. Also, it will be shown in the
proof of this theorem in the Appendix that conditions (b) and (c) of
Theorem~\ref{Suff_theorem} follow from condition
(\ref{dF.dF<0}) when $h_{a_1}^{x,s}(\cdot)$ and $h_{a_2}^{x,s}(\cdot)$
are smooth and strictly convex. On the other hand, condition (\ref{dF=dF=0}) was ruled out in
Theorem~\ref{Suff_theorem}; see condition (c) in that 
theorem. Indeed, conditions (\ref{dF.dF<0}) and (\ref{dF=dF=0}) are
mutually exclusive. 

\subsection{Rewarding certain actions leads to randomized jamming
  strategies}\label{main.3}

Attacker's decisions to pass/block transmission of control
information may be based on considerations other than cost to the
controller. These considerations encoded in the
jamming cost $g^0(a,s)$ in (\ref{sigma}) may either
discourage the jammer from launching an attack, or conversely encourage it
to undertake a  denial-of-service attack. The cost of link switching
may be one consideration behind the jammer's decisions whether to 
launch an attack. Also, the jammer
may be rewarded for being inactive to ensure it is not detected; e.g., this situation may occur when the
controller monitors the 
communication link, and an anomaly in the link behaviour can reveal the
jammer. In this section we show that when the attacker is rewarded for undertaking certain actions, a set $\mathcal{G}$
considered in Lemma~\ref{suff_cond} and Theorem~\ref{Suff_theorem} arises
naturally.  

Most generally, the reward scenario can be captured by reserving
special actions in the jammer's action space $\mathcal{A}$; such actions
will typically attract a distinctly different cost.   
For simplicity let $a_\circ$ be the only reserved action. Next, suppose the
remaining actions are ranked in accordance with 
their contribution to the game payoff. 

\begin{assumption}\label{sigma.A2}
For any two actions $j,k\in\mathcal{A}$, $j,k\neq a_\circ$, if $j<k$ then
for all $ u\in \overset{\circ}U(x)$
\begin{eqnarray}
  \label{chan.ineq}
  \mathbb{E}^{u,p}[\Sigma(x^+,u,a)|a=j]>
  \mathbb{E}^{u,p}[\Sigma(x^+,u,a)|a=k], 
\end{eqnarray}
here $\overset{\circ}U(x)$ is the interior of the compact set $U(x)$ from
Theorem~\ref{BO.T4}.
\end{assumption}

Under Assumption~\ref{sigma.A2}, the jammer who seeks a higher value of
the game will favour actions with lower numbers,
since these actions generate larger payoff. In contrast, the
controller should be forcing the jammer into
utilizing actions with higher numbers. The reserved action $a_\circ$ has
been excluded from ranking. Doing so is instrumental to provide the jammer
with a genuine choice between taking the reward and blocking/passing
transmission. 

We now show that using the action
ranking from Assumption~\ref{sigma.A2}, the value and the saddle
points of the game (\ref{inf_sup}) can be characterized by solving a game
over a reduced jammer's action space $\mathcal{G}$ consisting of the reserved action $a_\circ$ and one of the remaining actions which delivers the highest 
payoff to the jammer when it seeks to block communications between the
controller and the plant.   

Note that when $a_\circ\neq 1$, then according to (\ref{chan.ineq}), the
jammer's highest payoff among `regular' (i.e., not reserved) actions is
associated with action $a=1$. Alternatively, if 
$a_\circ= 1$, then according to (\ref{chan.ineq}), the highest
payoff among the actions from $\mathcal{A}\backslash \{a_\circ\}$ is
delivered when the jammer selects $a=2$. In both cases, the reduced action
space will contain only two actions from 
$\mathcal{A}$; in the first case, $\mathcal{G}=\{1,a_\circ\}$, or
$\mathcal{G}=\{1,2\}$ in the second case. The analysis of the reduced game
is the same in 
both cases, therefore we only consider the case
$\mathcal{G}=\{1,a_\circ\}$. In Section~\ref{Examples}, detailed examples
will be given to illustrate this case.   

Now that the set $\mathcal{G}$ has been established, the existence of
non-pure strategies in the ANC game with reserved actions can be derived
from the results in the previous section. The first result is a corollary
from Theorem~\ref{Suff_theorem} and is applicable when the
control input is a vector. 

\begin{corollary}\label{Suff_theorem.2.a}
Suppose Assumptions~\ref{sigma.A3} and~\ref{sigma.A2} hold. Furthermore,
suppose that for given initial $x$, $s$ there exists $u^*=u^*(x,s)
\in \overset{\circ}{U}$ such that the functions $h_1^{x,s}(\cdot)$,
$h_{a_\circ}^{x,s}(\cdot)$  satisfy the conditions
\begin{enumerate}[(a)]
\item $h_1^{x,s}(u^*)=h_{a_\circ}^{x,s}(u^*)$;
\item For all $u\neq u^*$, $h_1^{x,s}(u)>h_1^{x,s}(u^*)$
  or $h_{a_\circ}^{x,s}(u)> h_{a_\circ}^{x,s}(u^*)$;
\item $u^*$ is not a minimum of $h_1^{x,s}(\cdot)$,
  $h_{a_\circ}^{x,s}(\cdot)$.  
\end{enumerate}
Then, the zero-sun game (\ref{inf_sup}), (\ref{sup_inf}) admits a
non-pure saddle point $(u^*,p^*)$ with non-pure vector $p^*$ supported on
$\mathcal{G}=\{1,a_\circ\}$. 
\end{corollary}

\emph{Proof: }
  Under Assumption~\ref{sigma.A2}, $h_1^{x,s}(u^*)> h_k^{x,s}(u^*)$
  $\forall k\not\in\{1,a_\circ\}$. Together with condition (a), 
  this observation verifies condition (a) of
  Lemma~\ref{suff_cond}, with $\mathcal{G}=\{1,a_\circ\}$.  Conditions (b)
  and (c) of that lemma trivially follow from (b) and (c) in this
  corollary. Thus, from Lemma~\ref{suff_cond}, the ANC game has a saddle
  point $(u^*, p^*)$ with $ p^*$ 
  supported on $I( p^* ) \subseteq \mathcal{G}$. Since $|\mathcal{G}|=2$,
  and $p^*$ is not pure, then $|I(p^* )|=2$, i.e, $0<
  p^*_1<1$ and $0<p^*_{a_\circ}<1$. 
\hfill$\Box$

The second corollary follows from Theorem~\ref{saddle.point} and
applies when the control input is scalar. It eliminates condition
(\ref{FF>F}) which is the gap between the necessity and sufficiency
statements in that theorem.

\begin{corollary}\label{saddle.point.a}
Suppose Assumptions~\ref{sigma.A1'}, \ref{sigma.A2}
hold. The zero-sum game (\ref{inf_sup})
admits a non-pure saddle point $(u^*, p^*)$, $u^*\not\in U^d$, with $p^*$
supported on $\mathcal{G}=\{1,a_\circ\}$, if and
only if there exists $\bar u\in
\overset{\circ}U(x)\backslash U^d$ satisfying (\ref{F=F}) and either
(\ref{dF.dF<0}) or (\ref{dF=dF=0}) with $a_1=1$ and $a_2=a_\circ$.
\end{corollary}

\emph{Proof: } 
The corollary directly follows from Theorem~\ref{saddle.point} since
condition (\ref{FF>F}) of the sufficiency part of that theorem 
is trivially satisfied under Assumption~\ref{sigma.A2}.

\subsection{Remarks on the single-stage payoff}\label{Remark-on-signle}
The foregoing analysis of the single-stage ANC
game (\ref{inf_sup}) has relied on the properties of conditional payoff
functions  $h_i^{x,s}(\cdot)=\mathbb{E}^{u,p}[\Sigma(x^+,\cdot,a)|a=i]$
which in turn follow from coercivity, convexity and
continuity of $\Sigma(F(x,\cdot),\cdot,a)$ and $\Sigma(F(x,0),\cdot,a)$. Of
course these properties can be readily validated from the corresponding
properties of $\sigma^0$ and $\sigma^1$ through (\ref{sigma}).  However,
we stress that the results in this section are in fact
more general in that they apply to functions $\Sigma$ more general than
(\ref{sigma}); this observation will be useful in the next section.   

\section{Linear-quadratic controller-jammer games}\label{Examples}

In this section, we specialize the results of
Section~\ref{Static.ANC.Section} to one-stage controller-jammer ANC games
where the plant (\ref{plant.t}) is linear, and hence (\ref{plant}) becomes 
\begin{equation}
\label{plant.lin}
x^+ = A x + b Bu,
\end{equation}
and the cost of control  is quadratic. Also, we give examples of
the situations alluded in Section~\ref{Static.ANC.Section}, where
rewarding certain actions leads to randomized jamming.   

Two such games will be considered. In the first game the jammer is rewarded
for remaining stealthy, while in the second game its decisions are
determined by the premium the controller must pay for
terminating the game
(e.g., as a cost of repair or cost of re-routing control signals). We show
that in both games, there is a region in the plant state space where        
the jammer's optimal policy is to randomize among its actions, and an
optimal control response to this jamming policy is nonlinear. We
will also revisit the game from~\cite{Abhi} and will show that the
deterministically predictable behaviour of the jammer observed in~\cite{Abhi} can be predicted using our results.  

In order to provide a clear context of the attack strategies
resulting from these games we revisit the intelligent DoS by re-routing;
see Section~\ref{relation}. That is, we assume that 
the jammer's actions are to randomly select one of the available communication
channels, i.e., $\mathcal{A}=\mathcal{F}$, and $P(a)$ is defined as
in (\ref{Pt.1}). In addition, we will assume that  $u\in\mathbb{R}^1$;
this will allow us to apply Theorem~\ref{saddle.point} and, under
 Assumption~\ref{sigma.A2}, Corollary~\ref{saddle.point.a}. 

Thanks to the linear-quadratic nature of the games, in all three problems
the conditional 
expected payoff (\ref{hj}), given the jammer's action $a=j$, will have the
form\footnote{The superscript $^{x,s}$ is suppressed in this section, for
  economy of notation.}   
\begin{eqnarray}
h_j(u)
      &=&\gamma_j(x)+u^2+r_jq_ju(u+2\beta(x)),  
\label{hj.LaU3}
\end{eqnarray}
where $r_j\ge 0$ is a constant, and $\gamma_j(x)\ge 0$ for all $j$. Also, all
available channels are assumed to be ordered according to their probability
to become passing, that is, the probabilities (\ref{q_j}) are assumed to
be ordered as    
\begin{eqnarray}
q_1<q_2<\ldots<q_n.
\label{qq}
\end{eqnarray}

\begin{lemma}\label{U.LQ}
The set $U(x)=\{u:~u(u+2\beta(x)) \le 0 \}$  verifies properties
stated in Theorem~\ref{BO.T4}. Also, let $a_\circ\neq 1$ denote a reserved
action, and suppose $r_j=r>0$  and $\gamma_j(x)=\gamma(x)$ for all
$j\neq a_\circ$. Then under condition (\ref{qq}), 
Assumption~\ref{sigma.A2} is also satisfied with $U(x)$ defined above. 
\end{lemma}

\emph{Proof: }
We fix $x$ and assume $\beta(x)>0$. The case $\beta(x)<0$ can be analyzed
in a similar manner, while the case $\beta(x)=0$ is trivial.

The functions $h_j(u)$ defined in (\ref{hj.LaU3}) are continuous and strictly
convex in $u$. They are monotone decreasing on the interval
$(-\infty,-2\beta(x)]$ and are monotone increasing on the interval
$[0,+\infty)$. 
Therefore, $\hat h(u)=\max_j h_j(u)$ and 
$\bar h(u)=\sum_j p_jh_j(u)$ are 
also monotone decreasing on the interval
$(-\infty,-2\beta(x)]$ and are monotone increasing on $[0,+\infty)$. Since
both functions are continuous,
\begin{eqnarray*}
&& \min_{u\le -2\beta(x)}\hat  h(u) = \hat h(-2\beta(x)), \qquad \min_{u\ge 0}\hat  h(u) = \hat h(0), \\
&& \min_{u\le -2\beta(x)} \bar h(u) = \bar h(-2\beta(x)),
\qquad \min_{u\ge 0} \bar h(u) = \bar h(0). 
\end{eqnarray*}
Therefore, since the set $U(x)$ is closed and contains the points $0$ and
$-2\beta(x)$, then
\begin{eqnarray*}
\lefteqn{\inf_u \sup_{p\in\mathcal{S}_{N-1}} \bar h(u)= \inf_u \hat h(u)} && \\ 
&=&\min\left[ \inf_{u\le -2\beta(x)} \hat h(u), \inf_{u\in U(x)}\hat h(u),
  \inf_{u\ge 0}\hat  h(u)\right]  \\
 &=& \inf_{u\in U(x)}\hat h(u)=\inf_{u\in U(x)}\sup_{p\in\mathcal{S}_{N-1}} \bar
 h(u). 
\end{eqnarray*}
Also, for every $p\in \mathcal{S}_{N-1}$, $\inf_u \bar h(u)$ is attained at
$u=-\frac{\sum_j p_jr_jq_j}{1+\sum_j p_jr_jq_j}\beta(x)\in U(x)$,
therefore $\displaystyle \inf_u \bar h(u)= \inf_{u\in U(x)}\bar h(u)$ and  
$
\displaystyle\sup_{p\in\mathcal{S}_{N-1}}\inf_u \bar h(u)=
\sup_{p\in\mathcal{S}_{N-1}}  \inf_{u\in U(x)}\bar h(u). 
$
This proves that an optimal response of the minimizing player in the game
(\ref{inf_sup}) lies within the interval $[-2\beta(x),0]$. 

To verify the second claim of the lemma, we note that for any $j,k\neq
a_\circ$ if $j>k$ then due to (\ref{qq}) $h_j(u)<h_k(u)$
for all $u\in (-2\beta(x),0)$ i.e., (\ref{chan.ineq}) is satisfied.
\hfill$\Box$

\subsection{LQ control under reward for
  stealthiness}\label{LQR.reward.section} 
 
Consider a controller-jammer game for the plant (\ref{plant.lin})
with initial conditions $x^0=x$ and $s^{0}=s$ and 
the quadratic payoff (\ref{sigma}), with
$\sigma^0(x,u)=\|x\|^2+u^2$, $\sigma^1(x^+)=\|x^+\|^2$, 
$g^0(a,s)=-\tau \delta_{a,s}$.
Here, 
$\tau>0$ is the constant payoff which the jammer receives if 
it does not re-route control packets through a different channel. The
rationale here is to reward the jammer for not switching channels when
excessive switching may reveal its presence 
(hence rewarding  stealthiness), or may drain its resources. Thus, the
reserved action is to maintain transmission through the initial channel
$s$,  i.e., $a_\circ=s$.  
%
%
The corresponding function $\Sigma$ (\ref{sigma}) in this case is
\begin{equation}
\Sigma(x^+,u,a)=\begin{cases} \|x\|^2+u^2+\|x^+\|^2, & a\neq s, \\
\|x\|^2+u^2+\|x^+\|^2+\tau,  & a=s,
\end{cases}
\label{LQ.cost.sigma}
\end{equation}
and the static LQ ANC game is to find a control strategy $u^*$
and a jammer's non-pure strategy $p^*$ which form a saddle point of the game
(\ref{inf_sup}) for the plant (\ref{plant.lin}), with the payoff
(\ref{LQ.cost.sigma}). We now show that Corollary~\ref{saddle.point.a} can be
used for that. Indeed, the functions
$h_j(u)=\mathbb{E}^{u,p}[\Sigma(x^+,u,a)|a=j]$ have the form 
(\ref{hj.LaU3}), with $\gamma_j(x)= x'(I+A'A)x$, $j\neq s$
$\gamma_s(x)=x'(I+A'A)x+\tau$, 
and $\beta(x)=\frac{1}{\|B\|^2}B'Ax$, $r_j=\|B\|^2$ for all
$j\in\mathcal{A}$. 
According to Lemma~\ref{U.LQ}, with $a_\circ=s$, 
Assumption~\ref{sigma.A2} is satisfied in this special case. Furthermore, 
Assumption~\ref{sigma.A1'} is also satisfied, due to linearity of the plant
and a quadratic nature of the payoff. Then the analysis of the ANC game can
be reduced to verifying whether the payoff functions 
for the reduced zero-sum game, namely 
$h_1^{x,s}(u)= h_1(u)$ and $h_{a_\circ}^{x,s}(u)= h_s(u)$
satisfy the conditions of Corollary~\ref{saddle.point.a}.     

Using (\ref{hj.LaU3}), condition (\ref{F=F}) reduces to the
equation to be solved for $\bar u\in (-2\beta(x),0)$,
\begin{equation}
  \label{F=F.LQ}
  \bar u \left(\bar u + \frac{2}{\|B\|^2}B'Ax\right)=
  \frac{\tau}{\|B\|^2(q_1-q_s)}, 
\end{equation}
which admits real solutions if $\frac{1}{\|B\|^2}x'A'BB'Ax\ge
\frac{\tau}{q_s-q_1}$.    
Also, condition (\ref{dF.dF<0}) reduces to the condition
\begin{equation}
  \label{dF.dF<0.LQ}
  -\frac{\|B\|^2q_s}{1+\|B\|^2q_s}< \bar u <
  -\frac{\|B\|^2q_1}{1+\|B\|^2q_1}. 
\end{equation}
The analysis of conditions (\ref{F=F.LQ}), (\ref{dF.dF<0.LQ}) shows that only
one of the solutions of equation (\ref{F=F.LQ}), namely
\begin{eqnarray}
\bar u&\triangleq &u^*\!=\! -\frac{1}{\|B\|^2}B'Ax\!\left(\!1-\!\sqrt{\frac{\tau
    \|B\|^2}{(q_1-q_s)x'A'BB'Ax}}\!\right) \quad
\label{u*}
\end{eqnarray}
satisfies (\ref{dF.dF<0.LQ}) provided
\begin{eqnarray}
&&R_1< x'A'BB'Ax <R_s, 
\label{z.LaU3a} \\
&& 
R_1\triangleq \frac{(1+\|B\|^2q_1)^2}{(1+\|B\|^2q_1)^2-1}\frac{q_s-q_1}{\tau
  \|B\|^2}, \nonumber \\
&& 
R_s\triangleq \frac{(1+\|B\|^2q_s)^2}{(1+\|B\|^2q_s)^2-1}\frac{q_s-q_1}{\tau
  \|B\|^2}. \nonumber 
\end{eqnarray}
Condition (\ref{z.LaU3a}) describes the region in the state space in which the
jammer's optimal policy is to choose randomly between the initial channel
$s$ and the most blocking channel $1$. In the
case where the plant (\ref{plant.lin}) is scalar and $B=1$, we recover
exactly the condition obtained in~\cite{LaU3a} by direct
computation. That is, Corollary~\ref{saddle.point} confirms the existence of
the jammer's non-pure optimal strategy for this region. We refer the
reader to~\cite{LaU3a} for the exact value of the optimal vector $p^*$;
the calculation for the multidimensional plant (\ref{plant.lin}) follows
same lines, and is omitted for the sake of brevity. We
also point out that the optimal controller's policy (\ref{u*}) is
nonlinear. 

Let us now compare quadratic performance
of the optimal response control $u^*$ with performance of the
optimal guaranteed cost control law designed to control the plant
(\ref{plant.lin}) over a \emph{bona fide} packet dropping link with unknown
probability distribution of packet dropouts. Since the controller is unaware
of the attack and is also unaware of the precise statistical characteristics of
the channel, such a control strategy is 
a natural choice. Let $d$ denote the unknown probability distribution of
the transmission state of a packet dropping link perceived by the
controller. The set of all feasible probability distributions $d$ is
denoted $D$. The optimal 
guaranteed performance control $u^{\textrm{LQR}}$ is characterized by the condition  
\begin{eqnarray}
\lefteqn{\sup_{d\in D}
\mathbb{E}^d[\|x\|^2+|u^{\textrm{LQR}}|^2+\|x^+\|^2]} && \nonumber \\
&=&\inf_u\sup_{d\in D} \mathbb{E}^d[\|x\|^2+|u|^2+\|x^+\|^2]; 
  \label{lqr.opt}
\end{eqnarray}
here $\mathbb{E}^d$ denotes the expectation with
respect to the probability vector $d$. In particular, when the link is
optimally controlled by the 
jammer, we have $d^*=[(p^*)'q,~(1-(p^*)'q)]'$. Hence when $d^*\in D$, then 
\begin{eqnarray}
  \lefteqn{\mathbb{E}^{u^{\textrm{LQR}},p^*}[\|x\|^2+|u^{\textrm{LQR}}|^2+\|x^+\|^2]} &&
  \nonumber \\
&\le& \sup_{d\in D} \mathbb{E}^d[\|x\|^2+|u^{\textrm{LQR}}|^2+\|x^+\|^2] \nonumber \\
&=&\inf_u\sup_{d\in D} \mathbb{E}^d[\|x\|^2+|u|^2+\|x^+\|^2]. 
  \label{lqr.saddle.point.2}
\end{eqnarray}
On the other hand, since $(u^*,p^*)$ is the saddle point of
the one-stage min-max ANC game  which we have shown to admit
the upper value, we conclude  that for all $u$,
\begin{eqnarray}
\lefteqn{\mathbb{E}^{u^*,p^*}[\|x\|^2+|u^*|^2+\|x^+\|^2+\delta_{a,s}\tau]} &&
\nonumber \\ 
&\le&  \mathbb{E}^{u,p^*}[\|x\|^2+|u|^2+\|x^+\|^2+\delta_{a,s}\tau].
  \label{lqr.saddle.point}
\end{eqnarray}
Drop the term
$\mathbb{E}^{u^*,p^*}[\delta_{a,s}\tau]
=\mathbb{E}^{u,p^*}[\delta_{a,s}\tau]=p_s^*\tau$
on both sides of 
(\ref{lqr.saddle.point}),  
it then follows from (\ref{lqr.saddle.point.2}),
(\ref{lqr.saddle.point})
\begin{eqnarray}
  \lefteqn{\mathbb{E}^{u^*,p^*}[\|x\|^2+|u^*|^2+\|x^+\|^2]} &&
  \nonumber \\
&\le& 
\inf_u\sup_{d\in D} \mathbb{E}^d[\|x\|^2+|u|^2+\|x^+\|^2]. 
  \label{lqr.saddle.point.3}
\end{eqnarray}
This argument shows that when the jammer exercises its optimal randomized
strategy, unbeknownst to the controller, and forces the latter to 
perceive the communication link as a benign uncertain packet dropping
channel, the guaranteed cost control law designed under this 
assumption will have an inferior performance, compared with $u^*$, provided
$d^*\in D$. We interpret this situation as a signature of a successful DoS
attack by the jammer.   

From the above analysis, it appears that a possible line of 
defence against the DoS attack analyzed in this section could be 
controlling the system so as to avoid the
region defined by (\ref{z.LaU3a}). In
this case, the jammer will be forced to deterministically route
control packets over the genuine, albeit most damaging, packet dropping
channel. An impact of this defense on the system performance is an open
problem which will be studied in future work.

\subsection{Linear quadratic game with cost on loss of
  control}
We again consider the plant (\ref{plant.lin}), but with the  
 payoff
\begin{equation*}
\|x\|^2+v^2+\|x^+\|^2
=\|x\|^2+b^2u^2+\|x^+\|^2.
\label{LQ.cost.Abhi}
\end{equation*}
This is a one-step
version of the payoff considered in~\cite{Abhi}. The
payoff in this game directly depends on whether the control input is blocked.  
In this case, the functions $h_j$ have a form slightly different from those
in Section~\ref{LQR.reward.section}:
\begin{eqnarray*}
h_j(u)&=&x'(I+A'A)x \\
&& +q_j(1+\|B\|^2)u\left(u+\frac{2}{1+\|B\|^2}B'Ax\right).
\end{eqnarray*}
According to Lemma~\ref{U.LQ}, Assumption~\ref{sigma.A2} is still
satisfied (with the reserved
action $a_\circ=s$), and we can attempt to apply
Corollary~\ref{saddle.point.a}. For 
every $x\neq 0$, two points solve condition (\ref{F=F}),
$
\bar u=0$ and $\bar u=-\frac{2}{1+\|B\|^2}B'Ax$,
but none of them satisfy conditions (\ref{dF.dF<0}) or
(\ref{dF=dF=0}). Hence, the jammer's 
optimal strategy is to switch to the most blocking channel $1$, instead of
randomizing between $s$ and $1$. This finding is consistent with
the result  obtained in~\cite{Abhi}.  

\subsection{Linear-quadratic game with termination payoff} 
In this example, suppose the controller chooses a channel $s$ to transmit
information and receives messages from the plant as to which
channel was used for transmission. Once the controller detects change,
it terminates the game and pays a termination fee $T$, otherwise its terminal
cost is a regular state-dependent cost $\sigma^1(x^+)$. The jammer has a
choice between disrupting that channel  
$s$ and thus revealing itself, and holding off
the attack to remain undetected. The 
game payoff is then 
\[
\Sigma(x^+,u,a)=\|x\|^2+u^2+(1-\delta_{a,s})T + \delta_{a,s}\|x^+\|^2.
\]
Owing to the term $\delta_{a,s}\|x^+\|^2$, this payoff is not of the
form (\ref{sigma}), however its conditional expectation given $a=j$ 
has the form (\ref{hj.LaU3}), with $\gamma_j(x)=\|x\|^2+T$, $r_j=0$, $j\neq
s$, $\gamma_s(x)=x'(I+A'A)x$, $r_s=\|B\|^2$,
and $\beta(x)=\frac{1}{\|B\|^2}B'Ax$; it satisfies all the convexity and
coercivity conditions required in Section~\ref{Static.ANC.Section}, hence
the results developed in Section~\ref{Static.ANC.Section} can be applied in
this problem; see the remarks in Section~\ref{Remark-on-signle}. Without
loss of generality, we only consider the case where $B'Ax\le 0$. The
analysis of the case $B'Ax\ge  0$ follows the same lines. 

In the game considered, all the channels except 
$s$ have equal value for the jammer. Essentially, the jammer has to choose
between two actions: allow the controller to use its chosen channel $s$ or
reveal itself by selecting some other channel, e.g., channel 1. For that
reason, we select $\mathcal{G}=\{1,s\}$ and
proceed using Theorem~\ref{saddle.point}. Naturally, we assume $s\neq 1$,
since the case $s=1$ is trivial. 

According to our selection of the set $\mathcal{G}$, the functions
$h_{a_1}^{x,s}$ and $h_{a_2}^{x,s}$ in Theorem~\ref{saddle.point} are $h_1$ 
and $h_s$ from~(\ref{hj.LaU3}), respectively. It can be shown that if
$B'Ax\le 0$, $T\ge  x'A'(I-\frac{q_s}{\|B\|^2}BB')Ax$, then   
\begin{eqnarray}
\bar u=u^*&=&-\frac{1}{\|B\|^2}B'Ax \nonumber \\
&+&\sqrt{\frac{\|B\|^2T-x'A'(\|B\|^2I-(q_s)BB')Ax}{\|B\|^4q_s}} \quad
\label{baru.g2}
\end{eqnarray}
validates conditions (\ref{F=F}), (\ref{dF.dF<0}) of
Theorem~\ref{saddle.point} when 
\begin{equation}
\bar u<0 \quad \mbox{and} \quad (1+\|B\|^2q_s)\bar u + (q_s)x'A'B>0.
\label{int.x.2} 
\end{equation}
Substituting (\ref{baru.g2}) into (\ref{int.x.2}) yields
\begin{eqnarray}
&&x'\!A'\!Ax\!>\! T,~~ T\!>\!x'A'\!\left(\!I\!-\!\frac{(2+\|B\|^2q_s)q_s^2}{(1+\|B\|^2q_s)^2}BB'\!\right)\! Ax.\qquad
\label{int.x.3} 
\end{eqnarray}
Note that 
$
I-\frac{(2+\|B\|^2q_s)q_s^2}{(1+\|B\|^2q_s)^2}BB'>I-\frac{q_s}{\|B\|^2}BB'>0.
$
Therefore the region in the plant's state space where the game has a
non-pure saddle point is the intersection of the interior of the ellipsoid
defined by condition (\ref{int.x.3}) and the set $\{x:~ \|Ax\|^2>
T\}$. Such an intersection is clearly not an empty set, since
$I-\frac{(2+q_s\|B\|^2)q_s^2}{(1+\|B\|^2q_s)^2}BB'<I$.
  
Once again, as in Section~\ref{LQR.reward.section},
our analysis discovers a region in the state space where the
jammer's optimal strategy is to act randomly. In fact, in this
problem the choice is between allowing transmission through the 
channel $s$ selected by the controller and switching to some other channel ---
no matter which channel is selected to replace $s$, the value of the game
is not affected by this selection. The jammer's optimal policy vector $p^*$
can be computed directly in this problem to be an arbitrary vector in
$\mathcal{S}_{N-1}$ with the $s$-th component being equal to 
\begin{eqnarray}
  \label{p*.T}
p_s^*&=&-\frac{1}{\|B\|^2q_s} \nonumber \\
&\times&\!\left(\!1-\!
\sqrt{\frac{(q_s)x'A'Ax}{\|B\|^2T-x'A'(\|B\|^2I-(q_s)BB')Ax}}\!\right). 
\end{eqnarray}
Again, as
in Section~\ref{LQR.reward.section}, the controller's best response is
nonlinear in $x$.

\section{Existence of saddle-point strategies for multistage finite horizon
  ANC games}\label{multistage}

In this section we extend some of our previous results about the
existence of saddle-point strategies to the case of the multistage ANC game
(\ref{equi-point.t}). The game was posed in Section~\ref{sec_mod}, but our
generalization 
will be concerned with a special case where the function $F_t$ is a linear
function, $F_t(x,v)=A_tx+B_tv$, 
where $A_t$, $B_t$ are matrices of matching dimensions.  
That is, we restrict attention to linear systems of the form
\begin{equation} 
  \label{plant.t.linear}
x^{t+1}=A_tx^t+b^tB_tu^t.  
\end{equation}

\begin{assumption}\label{CL.assum}
\begin{enumerate}[(i)]
\item
For every $t=0,\ldots,T-1$, there exist scalars $e_t,d_t$ and functions $\alpha_t, \beta_t: \mathbb{R}_+ \rightarrow \mathbb{R}$,  with $\lim_{y \rightarrow +\infty}  \alpha_t(y) = \lim_{y \rightarrow +\infty}  \beta_t(y) = +\infty $ and $\alpha_t(y) \geq e_t$,  $\beta_t(y) \geq d_t$ such that 
\[
\sigma^t(u,x) \geq  \alpha_t(\|u\|) + \beta_t(\|x\|) \quad \forall x \in
\mathbb{R}^n, u \in \mathbb{R}^m. 
\]

\item
For all $t=0,\ldots,T-1$, the function $\sigma^t: \mathbb{R}^n \times
\mathbb{R}^m \rightarrow \mathbb{R}_+$ is convex in both its arguments. 

\item
The function $\sigma^T: \mathbb{R}^n \rightarrow \mathbb{R}_+$ is convex
and there exist a scalar $d_T$ and a function $\beta_T: \mathbb{R}_+
\rightarrow \mathbb{R}_+$,  with $ \lim_{y \rightarrow +\infty}  \beta_T(y)
= +\infty $ and  $\beta_T(y) \geq d_T$ such that 
\[
\sigma^T(x) \geq \beta_T(\|x\|).
\]
\end{enumerate}
\end{assumption}

\begin{assumption}\label{contin}
The function $\sigma^T(\cdot)$ is continuous. Furthermore, for every
$t=0,\ldots,T-1$, the function $\sigma^t(\cdot,u)$ is continuous uniformly in
$u\in \mathbb{R}^m$. Also, the functions $\sigma^t(x,\cdot)$
are continuous for every $x\in \mathbb{R}^n$. 
\end{assumption}

The main result of this section is now presented.

\begin{theorem}
\label{exist_theo}
Suppose Assumptions~\ref{CL.assum} and~\ref{contin}
hold. Then the 
value functions 
$\{V_t\}_{t=0}^T$ of the ANC game (\ref{equi-point.t}) 
defined recursively by 
\begin{eqnarray}
V_T(x,s) &=& \sigma^T(x) - g^T(s), \nonumber \\
V_t(x,s) &=& \inf_{u \in \mathbb{R}^n} \sup_{p \in
  \mathcal{S}_{N-1}} p'   
\mathcal{V}_t(x,s,u) \label{infsup_t} \\
&= &  \sup_{p \in \mathcal{S}_{N-1}} \inf_{u \in \mathbb{R}^n} 
p'\mathcal{V}_t(x,s,u). \label{supinf_t}
\end{eqnarray}
for all $t=0,\ldots,T-1$ and $x\in\mathbb{R}^n$, 
$s\in\mathcal{F}$, where $\mathcal{V}_t$ is a function
  $\mathcal{X}\times \mathbb{R}^m\to
  \mathbb{R}^{N}$ with components
\begin{eqnarray}
\lefteqn{[\mathcal{V}_t(x,s,u)]_a\triangleq
\sigma^t(x,u) - g^t(a,s)}  && \nonumber \\
&&+ 
\mathbb{E}_t\!\left[V_{t+1}(x^{t+1},s^{t+1})|
x^t\!=\!x,  s^t\!=\!s, u^t\!=\!u, a^t \!=\! a\right], \qquad 
\label{calV}
\end{eqnarray}
are all well defined, convex and continuous in $x$. 
Furthermore, the strategy pair ($u^*,p^*$) defined by 
\begin{eqnarray}
(u^*)^t&=&\arg \inf_{u \in \mathbb{R}^n} \sup_{p \in \mathcal{S}_{N-1}} p' 
\mathcal{V}_t(x,s,u),  \\
(p^*)^t&=&  \arg \sup_{p \in \mathcal{S}_{N-1}}
  \inf_{u \in \mathbb{R}^n} p' 
\mathcal{V}_t(x,s,u),
\end{eqnarray}
is a saddle-point strategy of the ANC game
(\ref{equi-point.t}). 
\end{theorem}

Theorem~\ref{exist_theo} introduces the HJBI
equation for the ANC game~(\ref{equi-point.t}) associated with the linear
plant (\ref{plant.t.linear}), see (\ref{infsup_t}), (\ref{supinf_t}).
Such an equation reduces finding jammer's
optimal strategies and the corresponding controller's best responses for the
multi-step ANC problem to a one-step game-type problem of the type
considered in the previous sections. Unfortunately, even for linear plants it
appears to be difficult to obtain an analytical answer to the question
whether the jammer's corresponding optimal strategy is randomized or
not. The main difficulty appears to be obtaining a closed form expression
for $V_t(x,s)$  --- of course, the tools developed in the previous
sections can be used to characterize a possibility of non-pure equilibria
numerically. An
alternative is to consider greedy jamming strategies, where at each time
step the jammer pursues a strategy which is (sub)optimal only at this particular
time. For example, one possibility is to use a lower bound for $V_{t+1}$ in the
HJBI equation~(\ref{calV}),
\begin{eqnarray*}
&&\inf_{u \in \mathbb{R}^n} \sup_{p \in
  \mathcal{S}_{N-1}} p'
\tilde{\mathcal{V}}_t(x,s,u), \label{infsup_t.greedy} 
\\
\lefteqn{[\tilde{\mathcal{V}}_t(x,s,u)]_a\triangleq
\sigma^t(x,u) - g^t(a,s)}  && \nonumber \\
&&\quad + 
\mathbb{E}_t\left[\nu_{t+1}(\|A_tx+b^tB_tu\|)\big|
s^t=s, a^t = a\right]. 
\label{calV.greedy}
\end{eqnarray*}
While not optimal, such greedy strategies may 
allow the jammer launch a randomized DoS attack which will likely be as
difficult to detect as an optimal one. The effect of such an
attack on the overall system performance and defence strategies against
it are an open question.

\section{Discussion and conclusions}\label{Conclusions}

In this paper we have analyzed a class of control problems over
adversarial communication links, in which the jammer strategically
disrupts communications between the controller and the plant. Initially, we have
posed the problem as a static game, and have given necessary and sufficient
conditions for such a game to have a non-pure saddle point. This allows a characterization of a set of plant's initial states for
which a DoS attack can be mounted that requires a nontrivial controller's
response. 

For instance, in two linear quadratic
problems analyzed in Section~\ref{Examples} the optimal control law is
nonlinear. This gives the jammer an advantage over any linear control
policy in those problems. The jammer achieves this by randomizing
its choice of a packet-dropping transmission regime rather than direct jamming. In those problems,
the part 
of the state space where the jammer randomizes is determined by the
jammer's cost of switching (reward for not switching), cost of termination, and transition probabilities of the current and the most
blocking regimes. If these parameters can be 
predicted/estimated by the controller, it has a chance of mitigating the
attack by either eliminating those regions, or steering the plant so that
it avoids visiting those  regions.      

Also, a multi-stage finite-horizon game has been considered for linear plants.
We have shown that equilibria for that game can be found by solving
similar one-stage games, although in general it is difficult to
obtain closed form solutions for these 
games, and one may need to resort to solving them numerically. As an
alternative, a greedy suboptimal analysis has been proposed. 

Future work will be directed to further understanding conditions for
DoS attacks, with the aim to obtain a deeper insight into dynamic/multi-step ANC
problems. Another interesting question is whether associating a distinct
payoff with one of the channels is necessary for the jammer to resort to
randomization. The system closed-loop stability under the proposed randomized
  jamming attack is also an interesting problem, namely the question whether
  it is possible for the jammer to degrade control performance and avoid
  being caught due to causing an instability. Analysis of this 
  problem requires a different, infinite horizon problem formulation.

\section{Appendix}\label{Appendix}

\subsection{Proof of Proposition~\ref{Markov}}

Since the process $s^t$ is a Markov chain,
for each $t=0,\ldots, T-1$ and an arbitrary measurable set
$\Lambda\times S\subseteq \mathcal{X}$, 
\begin{eqnarray*}
\lefteqn{\mathrm{Pr}(x^{t+1}\in\Lambda,s^{t+1}\in S|\{(x^\theta,s^\theta),u^\theta,a^\theta\}_{\theta=0}^t)}&& \nonumber
\\
&&=
\sum_{i\in S} P_{s^t, i}^t(a^t)
\mathrm{Pr}(x^{t+1}\in\Lambda|s^{t+1}=i, \{(x^\theta,s^\theta),u^\theta,
a^\theta\}_{\theta=0}^t). \nonumber
\end{eqnarray*}
Furthermore, from (\ref{q_j}) and (\ref{plant.t}), since given $s^{t+1}$,
$b_t$ is conditionally independent of $\{(x^\theta,s^\theta),u^\theta,
a^\theta\}_{\theta=0}^t$, then
\begin{eqnarray}
  \label{markov.prob}
\lefteqn{\mathrm{Pr}(x^{t+1}\in\Lambda,s^{t+1}\in S|
  \{(x^\theta,s^\theta),\{u^\theta,a^\theta\}_{\theta=0}^t)} &&
  \nonumber \\
&&=
\sum_{i\in S}
\mathrm{Pr}(x^{t+1}\in\Lambda|s^{t+1}=i, x^t,u^t)P_{s^t,i}^t(a^t)
  \nonumber \\
&&=
\mathrm{Pr}(x^{t+1}\in\Lambda,s^{t+1}\in S|x^t,s^t,u^t,a^t),
\end{eqnarray}
proving that $\{x^t,s^t\}_{t=0}^T$ is a Markov process.

\subsection{Proof of Theorem~\ref{BO.T4}}

First, we observe that since the function
$h_a^{x,s}(\cdot)$ is continuous and coercive, then according
to~\cite[Theorem~1.4.1]{PSU-1988},  
$\inf_{u}h_a^{x,s}( u)>-\infty$. Let $\alpha_x$ be a constant such that
$\alpha_x \ge \max_{a,s} \inf_{u} h_a^{x,s}( u)$. Owing to the
coercivity property of $h_a^{x,s}(\cdot)$ there exists $M^{s}_a(x)$
such that if $\|u\|>M^{s}_a(x)$ then  
$
h_a^{x,s}(u) > \eta_a^{s,x}(\|u\|)> \alpha_x.
$
Let $M(x)=\max_{a,s}M^{s}_a(x)$, and consider the ball
$U(x)=\{u:\|u\|\le M(x)\}$. Then the following facts hold
\begin{enumerate}[(i)]
\item
For all $s$,
$U^{s}(x)\triangleq\{u: \max_{a\in\mathcal{A}} h_a^{x,s}(u)\le \alpha_x\} \subseteq U(x).
\label{Uh.1}
$
Indeed, if $u\in U^{s}(x)$, then $h_a^{x,s}(u)\le \alpha_x$ therefore
$\|u\|\le M^{s}_a(x)\le M(x)$.

\item
Also, for all $p\in\mathcal{S}_{N-1}$ and $x\in\mathbb{R}^n$,
$s\in \mathcal{F}$,   
\begin{equation}
\{u: p'h^{x,s}(u) \le \alpha_x \}\subseteq U(x) \quad \forall
p\in\mathcal{S}_{N-1}. 
\label{Uh.2}
\end{equation}
Indeed, if $\|u\|> M(x)$, then $h_a^{x,s}(u) > \alpha_x$ for all
$a\in\mathcal{A}$ and we must have $p'h^{x,s}(u)>\alpha_x$. Therefore
$p'h^{x,s}(u) \le \alpha_x$ implies that $\|u\|\le M(x)$. 
\end{enumerate}

Using (i) and the fact that $p'h^{x,s}(u)$ is linear in $p$ and
therefore $\sup_{p\in \mathcal{S}_{N-1}}
p'h^{x,s}(u)=\max_{a\in \mathcal{A}} h^{x,s}_a(u)$, 
we obtain 
\begin{eqnarray}
J_1 &=& \inf_{u\in \mathbb{R}^m}\max_{a\in \mathcal{A}} h_a^{x,s}(u)  \le \inf_{u\in U(x)}\max_{a\in \mathcal{A}} h_a^{x,s}(u)  \nonumber \\
& \le& \inf_{u\in U^{s}(x)}\max_{a\in \mathcal{A}} h_a^{x,s}(u)= \inf_{u\in
  U^{s}(x)}\sup_{p\in \mathcal{S}_{N-1}} p'h^{x,s}(u). \qquad 
\label{Uh.3}
\end{eqnarray}
On the other hand, by definition, for $u\in\mathbb{R}^m\backslash
U^{s}(x)$, $\displaystyle\max_{a\in \mathcal{A}} h^{x,s}(u)> \alpha_x$, therefore
$\displaystyle
J_1\! =\!\!\inf_{u\in U^{s}(x)} \sup_{p\in \mathcal{S}_{N-1}}\!\!\!
  p'h^{x,s}(u).
$
Together with (\ref{Uh.3}) this yields~(\ref{inf_U_sup}).

In a similar manner, from (\ref{Uh.2}) it follows that $\forall p\in
\mathcal{S}_{N-1}$,
$
\inf_{u\in \mathbb{R}^m}  p'h^{x,s}(u) = \inf_{u\in U(x)} p'h^{x,s}(u)$.
The identity
$
J_2 = \sup_{p\in \mathcal{S}_{N-1}} \inf_{u\in U(x)} p'h^{x,s}(u)
$
immediately follows from that identity, i.e.,  the rightmost identity 
(\ref{J1J2}) holds.

We have established that both the upper value (\ref{inf_sup}) and the lower
value (\ref{sup_inf}) can be computed by performing minimization over the
compact set $U(x)$ which is also convex.
Since $\mathcal{S}_{N-1}$ is also compact and convex, and the
payoff function $p'h^{x,s}(u)$ is continuous and convex in $u$ 
for each $p\in \mathcal{S}_{N-1}$ and is continuous and concave
in $p$ for each $u \in U(x)$, the 
zero-sum game on the right-hand side of (\ref{inf_U_sup})
has a saddle point in $U(x)\times \mathcal{S}_{N-1}$,
but this saddle point may not be 
unique~\cite[Theorem 4, p.168]{B&O}. That is, the leftmost identity
(\ref{J1J2}) 
holds and the value of the zero-sum game over $U(x)\times
\mathcal{S}_{N-1}$ is finite.  Then, using (\ref{inf_U_sup}) and
(\ref{J1J2}) we conclude that both $J_1$ and $J_2$ are finite and equal.

\subsection{Proof of Lemma~\ref{suff_cond}}

The proof is a variation of well-known arguments regarding the connection
between convex duality and the existence of a saddle-point for the
Lagrangian, as well as  dual characterization of minimal elements of
convex sets (see, e.g., \cite{Boyd}). Because of our need to
incorporate unboundedness of vector $u^*$ and non-purity of vector $p$,
however, we find it clearer to provide a full
derivation here than to directly resort to these arguments. 

Consider the set 
\begin{eqnarray}
M:= \{z \in \mathbb{R}^{|\mathcal{G}|} \;|\; \exists u  \mbox{ such that }
H_\mathcal{G}(u) \leq z \};
\label{Mset}
\end{eqnarray}
here the notation $H_\mathcal{G}(u)$ refers to the vector comprised of the
components of $H(u)$ whose indexes belong to $\mathcal{G}$. Also, the
inequality in (\ref{Mset}) 
is understood component-wise. The interior and the boundary
of $M$ are denoted $\overset{\circ}{M}$ and $\partial M (=
M\setminus\overset{\circ}{M})$.   

$M$ is clearly convex since $H$ is.   In addition, assumption (b) implies
that $H_\mathcal{G}(u^*) \in \partial M$. Indeed, $H_\mathcal{G}(u^*)$
clearly belongs to $M$. If, in addition, we let $\mathbf{1}$ denote the
vector of all ones then, for any $r >0$ the point $z_r :=
H_\mathcal{G}(u^*) - \frac{r}{|\mathcal{G}|} \mathbf{1}$ belongs to the ball
$\mathcal{B}(H_\mathcal{G}(u^*),r)$ of center $H_\mathcal{G}(u^*)$ and
radius $r$ in $\mathbb{R}^{|\mathcal{G}|}$ but does \textit{not} belong to
$M$ since 
\begin{itemize}
\item $H_\mathcal{G}(u^*) > z_r$, and
\item according to (b), for any $u \neq u^*$, there exists $i$ such that
  $H_i(u) > H_i(u^*) >(z_r)_i$, i.e., $H_\mathcal{G}(u) \nleq z_r$ $\forall
  u\neq u^*$. 
\end{itemize}
 
We have thus shown that, for any $r>0$, $\mathcal{B}(H_\mathcal{G}(u^*),r)$
is \textit{not} a subset of $M$, which implies that $H_\mathcal{G}(u^*)$ is
not in the interior of $M$. Now, because $H_\mathcal{G}(u^*) \in \partial
M$, we can use the supporting hyperplane theorem (see, e.g., p.~51 of
\cite{Boyd}), to claim that there exists $\bar{p} \in
\mathbb{R}^{|\mathcal{G}|}$, $\bar{p}\neq 0$, such that 
\begin{equation}
\bar{p}' H_\mathcal{G}(u^*) \leq \bar{p}' z \mbox{ for all } z \in M. 
\label{separation}
\end{equation}
Now, we claim that $\bar{p} \in \mathbb{R}^{|\mathcal{G}|}_+$. Indeed, note
that if $z^0 \in M$, the 
ray $R_i := z^0 + \mathbb{R}^1_+ \mathbf{e}_i \subset M$, for any basis
vector $\mathbf{e}_i$. Hence, because of (\ref{separation}), the function
$z \mapsto 
\bar{p}' z$ is lower-bounded on $R_i$, which implies that $\bar{p}_i \geq
0$. Also note that $\bar{p}$ is non-pure, for otherwise, from
(\ref{separation}), there would exist $i \in \mathcal{G}$ such that  
$
H_i(u^*)= \bar{p}' H_\mathcal{G}(u^*)  \leq \bar{p}' H_\mathcal{G}(u) = 
H_i(u)$ $\mbox{ for all } u , 
$
(because $H_\mathcal{G}(u) \in M$ for all $u$), which contradicts (c).

Now, if we define $p^* \in \mathbb{R}^N$ by
$p^*_i=\frac{\bar{p}_i}{\bar{p}_1+\ldots+ \bar{p}_{|\mathcal{G}|}}$ for all $i \in \mathcal{G}$ and $p^*_i=0$ for all $i \notin \mathcal{G}$, we find that $p^* \in \mathcal{S}_{N-1}$, $I( p^*) \subset \mathcal{G}$ and
\begin{eqnarray*}
(p^*)' H(u^*) &=& \frac{1}{\bar{p}_1+\ldots+ \bar{p}_{|\mathcal{G}|}}
\bar{p}'H_\mathcal{G}(u^*) \\
&
\leq & \frac{1}{\bar{p}_1+\ldots +\bar{p}_{|\mathcal{G}|}} \bar{p}'H_\mathcal{G}(u) = (p^*)' H(u)
\end{eqnarray*}
$ \mbox{ for all } u$, which means that $u^*$ is a best response to $p^*$, i.e., 
$
u^* \in \arg\inf_{u \in \mathbb{R}^m} (p^*)'H(u).
$
Now, note that (a) implies that $p^*$ is a best response to $u^*$, i.e., $
p^* \in \arg\sup_{p \in \mathcal{S}_{N-1}} p'H(u^*),
$  since
\begin{align*}
(p^*)'H(u^*) &= \sum_{i \in \mathcal{G}} p^*_i H_i(u^*)=(\sum_{i \in \mathcal{G}} p^*_i) \max_{1 \leq j\leq N} H_j(u^*)\\
 &= \max_{1 \leq j\leq N} H_j(u^*) \geq p' H(u^*) \mbox{ for all } p \in \mathcal{S}_{N-1}.
\end{align*}
In fact, the same proof would show that any vector with support included in $\mathcal{G}$ is a best response to $u^*$. Now, 
\begin{eqnarray*}
 \inf_{u \in \mathbb{R}^m} \sup_{p \in \mathcal{S}_{N-1}} p'H(u) &\leq&
 \sup_{p \in \mathcal{S}_{N-1}} p'H(u^*)=(p^*)'H(u^*)  \\ 
= \inf_{u \in \mathbb{R}^m} (p^*)' H(u) 
&\leq & \sup_{p \in \mathcal{S}_{N-1}} \inf_{u \in \mathbb{R}^m} p' H(u),
\end{eqnarray*}
while it is always true that 
$
\sup_{p} \inf_{u} p' H(u) \leq  \inf_{u} \sup_{p} p' H(u).
$
This concludes the proof. 
\hfill$\Box$

\subsection{Proof of Theorem~\ref{saddle.point}}

\emph{Sufficiency. }
First we show that (\ref{F=F}), (\ref{FF>F}) and (\ref{dF.dF<0}) imply the
existence of a non-pure saddle point for (\ref{inf_sup}) supported on
$\mathcal{G}$. Without loss of generality suppose 
\begin{equation}\label{dF.dF<0.1}
\frac{dh_{a_1}^{x,s}(\bar u)}{d u}>0, \quad 
\frac{dh_{a_2}^{x,s}(\bar u)}{d u}<0.
\end{equation}
Since $h_{a_1}^{x,s}(\cdot)$ is convex, it follows from
(\ref{dF.dF<0.1}) that $\frac{dh_{a_1}^{x,s}(u)}{d u}$ is  non-decreasing
in the region $u\ge \bar u$. This is true for all points $u\ge \bar u$
including the points of the set $U^d$ where we have the inequality between
the right-hand side and left-hand side derivatives, 
$\frac{d^-h_{a_1}^{x,s}(u)}{d u}\le \frac{d^+h_{a_1}^{x,s}(u)}{d u}$. 
Hence $\frac{dh_{a_1}^{x,s}(u)}{d u}>0$ for all
$u\ge \bar u$ and thus $h_{a_1}^{x,s}(u)>h_{a_1}^{x,s}(\bar u)$ for all $u>\bar
u$. In the same manner we can show that  $h_{a_2}^{x,s}(u)<h_{a_2}^{x,s}(\bar u)$
for all $u<\bar u$. Also, $\bar u$ is not a minimum of 
$h_{a_1}^{x,s}( \cdot)$ and $h_{a_2}^{x,s}(\cdot)$
since $\frac{dh_{a_1}^{x,s}(\bar u)}{d u}\neq 0$, $\frac{dh_{a_2}^{x,s}(\bar
  u)}{d u}\neq 0$. The sufficiency of conditions  (\ref{F=F}), (\ref{FF>F}) and
(\ref{dF.dF<0}) now follows from Theorem~\ref{Suff_theorem}.

We now consider the second alternative case where (\ref{F=F}), (\ref{FF>F}) and
(\ref{dF=dF=0}) hold. Since $h_{a_1}^{x,s}(u)$ and $h_{a_2}^{x,s}(u)$ are
convex, (\ref{dF=dF=0}) implies that $\bar u$ is a global minimum
of both $h_{a_1}^{x,s}(u)$ and $h_{a_2}^{x,s}(u)$. Therefore for an
arbitrary $p\in \mathcal{S}_{N-1}$ with $p_a=0$ for $a\neq
\mathcal{G}$, 
$\bar u$ is a global minimum of 
$
\mathbb{E}^{u,p}\Sigma(x^+,u,a)=p'H(u),
$
where $H(u)=h^{x,s}(u)$ is the vector function defined in
Theorem~\ref{Suff_theorem}. That is, 
$
 \inf_u p' H(u)
=p' H(\bar u)
=h_{a_1}^{x,s}(\bar u)=h_{a_2}^{x,s}(\bar u).
$
It then follows that for arbitrary $\bar p, \tilde p \in
\mathcal{S}_{N-1}$ supported on $\mathcal{G}$
and $u\in\mathbb{R}^1$,
\begin{eqnarray}
\tilde p' H(\bar u)\le \bar p' H(\bar u)
\le \bar p'  H(u). 
\label{sp.1}
\end{eqnarray} 
The leftmost inequality holds since both expressions are equal to 
$h_{a_1}^{x,s}(\bar u)=h_{a_2}^{x,s}(\bar u)$.
 
Next, consider an arbitrary vector
$p\in\mathcal{S}_{N-1}$. From (\ref{FF>F}) we have  
\begin{eqnarray} 
p'H(\bar u)&=&\sum_{j\neq a_2}
p_j\mathbb{E}^{\bar u,p}[\Sigma(x^+,\bar u,a)|a=j] + p_{a_2}
h_{a_2}^{x,s}(\bar u) \nonumber \\
&\le& (\sum_{j\neq a_2} p_j)\mathbb{E}^{\bar u,p}[\Sigma(x^+,\bar u,a)|a=a_1] +
p_{a_2} h_{a_2}^{x,s}(\bar u)  \nonumber \\
&=& \tilde p' H(\bar u), 
\label{sp.3}
\end{eqnarray}
where $\tilde p$ is defined as $\tilde p_{a_1}=\sum_{j\neq a_2} p_j$,
$\tilde p_{a_2}=p_{a_2}$ and $\tilde p_a=0$ for $a\not\in \mathcal{G}$, and
is supported on $\mathcal{G}$. Then from  (\ref{sp.1}), it follows
that for all $u\in\mathbb{R}^1$, and $p\in\mathcal{S}_{N-1}$
$
p'H(\bar u)\le (\bar p)'H(\bar u)  \le (\bar p)'H(u). 
$
Hence for an arbitrary $\bar p \in
\mathcal{S}_{N-1}$ supported on $\mathcal{G}$, $(\bar
u,\bar p)$ is a saddle point of the game. Clearly, a non-pure $\bar p$
exists for this purpose.

\emph{Necessity. }
Let $(u^*,p^*)$, $u^*\not\in U^d$, 
be a non-pure saddle point of the game 
(\ref{inf_sup}) supported on $\mathcal{G}=\{a_1,a_2\}$.
That is, $0<p_{a_1}^*<1$ and
\begin{eqnarray}
 p' H(u^*)
\le (p^*)' H(u^*)
\le ( p^*)' H(u), \quad \nonumber \\ 
\forall
u\in\mathbb{R}^1,  p\in \mathcal{S}_{N-1}. 
\label{sp.*}
\end{eqnarray}
The last inequality implies
\begin{eqnarray*}
\inf_u \max_{p\in \mathcal{S}_{N-1}}p' H(u) && \le \max_{p\in
  \mathcal{S}_{N-1}} p' H(u^*) = (p^*)' H(u^*) \\
&& =  \inf_u (p^*)' H(u) 
\le  \max_{p\in \mathcal{S}_{N-1}}\inf_u p' H(u).
\end{eqnarray*}
However, the game has the value, therefore all the inequalities above are in
fact exact identities. That is,
 \begin{eqnarray*}
\inf_u \max_{p\in \mathcal{S}_{N-1}} p' H(u) 
= \max_{p\in \mathcal{S}_{N-1}}p' H(u^*). 
\end{eqnarray*} 
In other words, $u^*\in \arg \inf_u \max_{p\in
  \mathcal{S}_{N-1}} p' H(u)$, and taking into account 
linearity of the payoff function in $p$, we further have
$
u^*\in 
\arg \inf_u \max[h_{a_1}^{x,s}(u), h_{a_2}^{x,s}(u)].
$
We now show that $h_{a_1}^{x,s}(u^*)=h_{a_2}^{x,s}(u^*)$.
Suppose this is not true, and $h_{a_1}^{x,s}(u^*)<h_{a_2}^{x,s}(u^*)$. Then,
since $p_{a_1}^*>0$, $p_{a_2}^*>0$, and $p_{a_1}^*+p_{a_2}^*=1$  by
assumption,    
\begin{eqnarray*}
h_{a_2}^{x,s}(u^*)= (p_{a_1}^*+p_{a_2}^*) h_{a_2}^{x,s}(u^*) 
> (p^*)'H(u^*).
\end{eqnarray*}
The latter condition is in contradiction with the leftmost inequality in
(\ref{sp.*}). The converse inequality $h_{a_2}^{x,s}(u^*)<h_{a_1}^{x,s}(u^*)$ is also not possible, by the same argument. This proves
that 
$h_{a_1}^{x,s}(u^*)=h_{a_2}^{x,s}(u^*)$, i.e., $u^*$ satisfies the condition (\ref{F=F})
of the theorem.   

It remains to prove that $u^*$ satisfies ether (\ref{dF.dF<0}) or
(\ref{dF=dF=0}). We prove this by ruling out all other possibilities. 

\emph{Case 1: $\left(\frac{dh_{a_1}^{x,s}(u^*)}{d u}\right)
\left(\frac{dh_{a_2}^{x,s}(u^*)}{d u}\right)>0$.}

First suppose $\frac{dh_{a_1}^{x,s}(u^*)}{d u}<0$,
$\frac{dh_{a_2}^{x,s}(u^*)}{d u}<0$.  Then, in a
sufficiently small neighbourhood of $u^*$ one can find a point $u<u^*$ such
that 
\begin{eqnarray*}
h_{a_1}^{x,s}(u) < h_{a_1}^{x,s}(u^*) \quad \mbox{and} \quad
h_{a_2}^{x,s}(u) < h_{a_2}^{x,s}(u^*).
\end{eqnarray*}
That is, we found a point $u<u^*$ such that
$(p^*)' H(u) < ( p^*)' H(u^*)$.
This conclusion is in contradiction with the rightmost inequality
(\ref{sp.*}). The hypothesis that $\frac{dh_{a_1}^{x,s}(u^*)}{d
  u}>0$, $\frac{dh_{a_2}^{x,s}(u^*)}{d u}>0$ will lead to
a similar contradiction. These contradictions rule out Case 1.     

\emph{Case 2: Either $\frac{dh_{a_1}^{x,s}(u^*)}{d u}=0$
  and $\frac{dh_{a_2}^{x,s}(u^*)}{d u}\neq 0$, or
$\frac{dh_{a_1}^{x,s}(u^*)}{d u}\neq 0$
  and $\frac{dh_{a_2}^{x,s}(u^*)}{d u}= 0$.}

Suppose $\frac{dh_{a_1}^{x,s}(u^*)}{d u}=\beta>0$ and
$\frac{dh_{a_2}^{x,s}(u^*)}{d u}= 0$. 
Since $h_{a_1}^{x,s}(\cdot)$, $h_{a_2}^{x,s}(\cdot)$ are continuously
differentiable at $u^*$, then for any 
sufficiently small $\epsilon>0$, there exists $\delta>0$ such that for any
$u\in (u^*-\delta,u^*+\delta)$,
$
\frac{dh_{a_1}^{x,s}(u)}{d u}>\beta-\epsilon$, 
$\left|\frac{dh_{a_2}^{x,s}(u)}{d u}\right|<\epsilon. $
Let us choose $\epsilon$ so that $\epsilon< p_{a_1}^*\beta<\beta$ and 
consider the Taylor expansions of $h_{a_1}^{x,s}(u)$, $h_{a_2}^{x,s}(u)$, 
$u^*-\delta<u<u^*$, with the remainders in the Cauchy form   
\begin{eqnarray*}
h_{a_1}^{x,s}(u)= h_{a_1}^{x,s}(u^*)+\frac{d h_{a_1}^{x,s}(\xi_1)}{d u} (u-u^*),  \\  
h_{a_2}^{x,s}(u)= h_{a_2}^{x,s}(u^*)+\frac{dh_{a_2}^{x,s}(\xi_2)}{d u} (u-u^*), 
\end{eqnarray*} 
where $\xi_1,\xi_2\in (u,u^*)$. Since $h_{a_2}^{x,s}(u)$ is convex
and $\frac{dh_{a_2}^{x,s}(u^*)}{d u}= 0$, then
$-\epsilon<\frac{dh_{a_2}^{x,s}(\xi_2)}{d u}\le 0$ for
$u^*-\delta<u<\xi_2<u^*$. This leads us to conclude that
\begin{eqnarray*}
p_{a_1}^*\frac{dh_{a_1}^{x,s}(\xi_1)}{d u}
+p_{a_2}^*\frac{dh_{a_2}^{x,s}(\xi_2)}{d u} &>&
p_{a_1}^*(\beta-\epsilon) - p_{a_2}^*\epsilon >0.
\end{eqnarray*}
Then we have 
\begin{eqnarray}
(p^*)' H(u)&=&(p^*)' H(u^*) \nonumber \\
&+& \left( p_{a_1}^*\frac{dh_{a_1}^{x,s}(\xi_1)}{d u}
+ p_{a_2}^*\frac{dh_{a_2}^{x,s}(\xi_2)}{d
  u}\right)(u-u^*) \nonumber  \\
&<& (p^*)' H(u^*).
\end{eqnarray}
Again, we arrive at a contradiction with the assumption that $(u^*,p^*)$ is
a saddle point and must satisfy (\ref{sp.*}).  

Other similar possibilities in this case will lead to a contradiction as
well. This leaves two possibilities: either $u^*$ satisfies
(\ref{dF.dF<0}), or it satisfies (\ref{dF=dF=0}).

\subsection{Proof of Theorem~\ref{exist_theo}}\label{exist_theo.proof}

We need to show that each function $V_t$ is well-defined, i.e., that there
is indeed equality between (\ref{infsup_t}) and (\ref{supinf_t}) for any
$x,s$; this amounts to the zero-sum game solved at time $t$ having a
value. We will proceed by backwards induction and show that  the following
predicate holds for all $t=0,\ldots,T$: 
\begin{description}
\item[$(\mathbf{P}_t):$] $V_t(\cdot,s)$ is a well-defined continuous convex
  function for all $s\in \mathcal{F}$. In addition,
  there exist a function $\nu_t$ and scalar $v_t$ such that 
\begin{enumerate}[(a)]
\item
$\nu_t(y) \geq v_t$ for all $y$,
\item
$\lim_{y \rightarrow
  +\infty }\nu_t(y) = + \infty$ and, 
\item
$V_t(x,s) \geq \nu_t(\|x\|)$ for all
$x \in \mathbb{R}^n$, $s\in \mathcal{F}$.
\end{enumerate}
\end{description}

Clearly predicate $\mathbf{P}_T$ holds since $\sigma^T$ is convex and continuous and
according to 
Assumption~\ref{CL.assum}, 
 \begin{eqnarray}
V_T(x,s) &=& \sigma^T(x) - g^T(s) \ge  \beta_T(\|x\|)-\max_{s\in\mathcal{F}}g^T(s) \nonumber \\
&\triangleq & \nu_T(\|x\|) \ge v_T
\end{eqnarray}
where $v_T\triangleq d_T-\max_{s\in\mathcal{F}}g^T(s)$. 

Let us now assume that predicate $\mathbf{P}_{t+1}$ holds for $t \le T-1$
and show 
that $\mathbf{P}_t$ holds. Note that
\begin{eqnarray}
\lefteqn{[\mathcal{V}_t(x,s,u)]_a=
\sigma^t(x,u) - g^t(a,s)} && \nonumber \\
&& + \sum_{i=1}^{|\mathcal{F}|} P_{si}^t(a) \left[ 
(1-q_i^t)
V_{t+1}(A_tx,s) \right. \nonumber \\
&&  + \left. 
q_i^t
V_{t+1}(A_tx+B_tu,s) \right], 
\label{calV.1}
\end{eqnarray}
Using Assumption \ref{CL.assum} we then find that  for
all $a\in \mathcal{A}$, 
\begin{eqnarray}
\lefteqn{[\mathcal{V}_t(x,s,u)]_a
\geq  \alpha_t(\|x\|) + \beta_t(\|u\|) -
\max_{s \in \mathcal{F}} g_t(a,s)} && \nonumber  \\
&& \quad + \sum_{i=1}^{|\mathcal{F}|} P_{si}^t(a) \left[ (1-q^t_i)
\nu_{t+1}(\|A_tx\|)\right. \nonumber \\
&& \quad\qquad  + 
q^t_i \nu_{t+1}(\|A_tx+B_tu\|)  \nonumber \\
&& \quad \geq 
\alpha_t(\|x\|) + \beta_t(\|u\|) - \max_{s \in \mathcal{F}} g_t(a,s) +
v_{t+1}.
\label{minor}
\end{eqnarray}
That is, for all fixed $(x,s)\in\mathcal{X}$, $a\in \mathcal{A}$, the
function $[\mathcal{V}_t(x,c,s,\cdot)]_a$ is coercive on 
$\mathbb{R}^m$. 

Also, since according to predicate $\mathbf{P}_{t+1}$, $V_{t+1}(\cdot,s)$ is
a continuous function, then by Assumption~\ref{contin} the
function $[\mathcal{V}_t(x,s,\cdot)]_a$ is continuous on $\mathbb{R}^m$
for all $x\in \mathbb{R}^n$, $a\in \mathcal{A}$, and $s\in
\mathcal{F}$. 
Finally, we note that since $V_{t+1}(\cdot,s)$ is a convex function by
assumption, and $A_tx+B_tu$ is linear with respect to $u$, then
$V_{t+1}(A_tx+B_tu,s)$ is convex in $u$. Also, $\sigma^t(x,u)$ is convex in
$u$ by Assumption~\ref{CL.assum}. Hence we conclude that
$[\mathcal{V}_t(x,s,\cdot)]_a$ is convex.

We have verified all the conditions of Lemma~\ref{BO.T4},
which can now be applied to ascertain
that the $\inf\sup$ and $\sup\inf$ expressions in (\ref{infsup_t}) and
(\ref{supinf_t}) are equal and finite. Thus, the function $V_t(x,s)$ is
well-defined, and there exists a saddle point pair of strategies
$(u^*)^t(x,s)$, $(p^*)^t(x,s)$ defined by the static zero-sum game
(\ref{infsup_t}). 
Furthermore, it follows from (\ref{minor}) that $V_t(x,s)$
satisfies properties (a)-(c) stated in predicate $\mathbf{P}_t$ with  
the function $\nu_t(\cdot)$ and constant $v_t$ defined as    
\begin{eqnarray*}
&&\nu_t(y)=\alpha_t(y)+d_t
-\max_{a\in\mathcal{A}}\max_{s\in\mathcal{F}}g_t(a,s)+v_{t+1};
\\
&&v_t=e_t+d_t -\max_{a\in\mathcal{A}}\max_{s\in\mathcal{F}}g_t(a,s)+v_{t+1}.
\end{eqnarray*}

It remains to prove that $V_t(\cdot,s)$ is convex and continuous. 
For continuity, we note that composition of $V_{t+1}(\cdot,s)$
and $A_tx+B_tu$ is continuous uniformly in 
  $u$, since the latter function has this property and the former function
  is continuous. Then   
  $P_{si}^t(a)q_i^tV_{t+1}(A_tx+B_tu,s)$ is also continuous
  uniformly in $u$. Therefore, for every $\tilde x\in\mathbb{R}^n$ and
  $\epsilon>0$ 
one can find $\delta^{a,s}(\tilde x,\epsilon)>0$ which does not depend on
$u$ and such that $\|x-\tilde x\|< \delta^{a,s}(\tilde x,\epsilon)$ implies 
$
|(\mathcal{V}_t(x,s,u))_a
-(\mathcal{V}_t(\tilde x,s,u))_a|<\frac{\epsilon}{2}. 
$
Selecting $\delta(\tilde x,\epsilon)=\min_{a,s} \delta^{a,s}(\tilde
x,\epsilon)$ we obtain  
\begin{eqnarray*}
  p'\mathcal{V}_t(\tilde x,s,u)-\frac{\epsilon}{2}< p'\mathcal{V}_t(x,s,u) <
p'\mathcal{V}_t(\tilde x,s,u)+\frac{\epsilon}{2} \\
\forall x \in \{x: \|x-\tilde x\|\le \delta(\tilde x,\epsilon)\}, 
\  p\in\mathcal{S}_{N-1}.
\end{eqnarray*}
From here we readily conclude that
\begin{eqnarray*}
  \lefteqn{\inf_{u}\sup_{p\in\mathcal{S}_{N-1}} p'\mathcal{V}_t(\tilde
  x,s,u)- \epsilon} && \\
&&\le \inf_{u}\sup_{p\in\mathcal{S}_{N-1}} p'\mathcal{V}_t(x,s,u) 
<
  \inf_{u}\sup_{p\in\mathcal{S}_{N-1}} p'\mathcal{V}_t(\tilde
  x,s,u) + \epsilon,
\end{eqnarray*}
for all $x \in \{x: \|x-\tilde x\|\le \delta(\tilde x,\epsilon)\}$, proving
that $V_t(\cdot,s)$ is continuous at an arbitrarily chosen 
$\tilde x$. 

For convexity of $V_t(\cdot,s)$, we note that each function
$(x,u)\to (\mathcal{V}_t(x,s,u))_a$ is convex since $\sigma^t$ is convex
in $(x,u)$ and $V_{t+1}(\cdot,s)$ is convex by the induction
hypothesis. As a result, the function
$\sup_{p\in\mathcal{S}_{N-1}} p'\mathcal{V}_t(\tilde x,s,u)$
is convex in $(x,u)$; see \cite[Section~3.2.3]{Boyd}. In turn, using 
identity (\ref{inf_U_sup}) of Lemma~\ref{BO.T4},
we conclude that 
\begin{eqnarray*}
V_t(x,s)&=&\inf_{u\in\mathbb{R}^m}\sup_{p\in\mathcal{S}_{N-1}}
p'\mathcal{V}_t(\tilde x,s,u) \\
&=&\inf_{U(x)}\sup_{p\in\mathcal{S}_{N-1}}
p'\mathcal{V}_t(\tilde x,s,u)
\end{eqnarray*}
is convex since $U(x)$ is a convex set.

\end{document}